%% file: main.tex
\documentclass[pdftex,twocolumn,epjc3]{svjour3}          %
\RequirePackage[T1]{fontenc}
\smartqed  %
\RequirePackage{graphicx}
\RequirePackage{mathptmx}      %
\RequirePackage{flushend}
\RequirePackage[numbers,sort&compress]{natbib}
\RequirePackage{amsmath}
\RequirePackage[english]{babel} 
\RequirePackage{bm}              
\RequirePackage{lineno}    
\RequirePackage[latin9]{inputenc} 
\RequirePackage{seqsplit}
\usepackage{geometry}
\usepackage{citesort}
\usepackage{color}
\usepackage{booktabs}
\usepackage{amssymb}
\usepackage{lmodern}
\usepackage{epstopdf}
\usepackage{pxfonts}
\usepackage[table]{xcolor}
\definecolor{lightgray}{gray}{0.87}
\journalname{Eur. Phys. J. C}
\usepackage{widetext}
\RequirePackage[colorlinks,citecolor=blue,urlcolor=blue,linkcolor=blue]{hyperref}
\usepackage{ulem}
\usepackage[none]{hyphenat}
\usepackage{nicefrac}

\begin{document}
\sloppy
\title{Impact of the heavy quark matching scales in  PDF fits}
\author{The xFitter Developers' Team:
     V.~Bertone\thanksref{a,b}
\and D. Britzger\thanksref{c}
\and S.~Camarda\thanksref{d}  
\and A.~Cooper-Sarkar\thanksref{e}
\and A.~Geiser\thanksref{c}
\and F.~Giuli\thanksref{e}
\and A.~Glazov\thanksref{c}
\and E.~Godat\thanksref{f}
\and A.~Kusina\thanksref{g,h}
\and A.~Luszczak\thanksref{i}  
\and F.~Lyonnet\thanksref{f}
\and F.~Olness\thanksref{f}
\and R.~Pla\v cakyt\.e\thanksref{j}
\and V.~Radescu\thanksref{c,d} 
\and I.~Schienbein\thanksref{g}
\and O.~Zenaiev\thanksref{c}
}
\institute{Department of Physics and Astronomy,  VU University, NL-1081 HV Amsterdam, The Netherlands \label{a}
\and Nikhef Theory Group Science Park 105, 1098 XG Amsterdam, The Netherlands \label{b}
\and DESY Hamburg, Notkestra{\ss}e 85, D-22609, Hamburg, Germany \label{c}
\and CERN, CH-1211 Geneva 23, Switzerland \label{d}
\and University of Oxford,1 Keble Road, Oxford OX1 3NP, United Kingdom \label{e}
\and SMU Physics, Box 0175 Dallas, TX  75275-0175, United States of America \label{f}
\and Laboratoire de Physique Subatomique et de Cosmologie, Universit\'e Grenoble Alpes, CNRS/IN2P3,  
    \\ \null \qquad   53 avenue des Martyrs, 38026 Grenoble, France\label{g}
\and Institute of Nuclear Physics, Polish Academy of Sciences,  ul. Radzikowskiego 152, 31-342 Cracow, Poland \label{h}
\and  T.Kosciuszko Cracow University of Technology, 30-084 Cracow, Poland\label{i}
\and Institut f\"ur Theoretische Physik, Universit\"at Hamburg, Luruper Chaussee 149, D--22761 Hamburg, Germany \label{j}
}
\date{Received: date  / Accepted: date / Today:\today }
\authorrunning{The xFitter Developers' Team}
\titlerunning{Impact of the heavy quark matching scales in  PDF fits}
\maketitle
\begin{abstract}
  We investigate the impact of displaced heavy quark matching scales
  in a global fit.
  The heavy quark matching scale $\mu_{m}$ determines 
  at which  energy scale $\mu$ the  QCD
  theory transitions from $N_{F}$ to $N_{F}+1$ in the Variable Flavor
  Number Scheme (VFNS) for the evolution of the Parton Distribution
  Functions (PDFs) and strong coupling $\alpha_S(\mu)$.
  We study the variation of the matching scales, and their impact on a
  global PDF fit of the combined HERA data.
  As the choice of the matching scale $\mu_{m}$ effectively is a
  choice of scheme, this represents a theoretical uncertainty;
  ideally, we would like to see minimal dependence on this parameter.
  For the transition across the charm quark (from $N_{F}=3$ to $4$),
  we find a large $\mu_m=\mu_{c}$ dependence of the global fit $\chi^2$ at NLO, but this is
  significantly reduced at NNLO.
  For the transition across the bottom quark (from $N_{F}=4$ to $5$),
  we have a reduced $\mu_{m}=\mu_b$ dependence of the $\chi^2$ at both NLO and NNLO as compared to the charm.
  This feature is now implemented in xFitter  2.0.0, an open source
QCD fit framework. 
\end{abstract}

\newpage

\tableofcontents{}

\input{figures.tex}
\input{tables.tex}
\input{text.tex}

\begin{acknowledgements}
  The authors would like to thank M.~Botje, J.~C.~Collins, and J.~Rojo
  for valuable discussions.
  V.~B and A.~G. are particularly grateful to A.~Mitov,
  A.~Papanastasiou, and M.~Ubiali for many stimulating discussions on
  the role of the bottom-quark threshold for bottom-initiated
  processes at the LHC.
  We acknowledge the hospitality of CERN, DESY, and Fermilab where a
  portion of this work was performed.
  We are grateful to the DESY IT department for their support of the
  xFitter developers.
  This work was also partially supported by the U.S.\ Department of
  Energy under Grant No.\ DE-SC0010129.
  V.~B.~is supported by an European Research Council Starting Grant
  ``PDF4BSM''.  A.~L.~is supported by the Polish Ministry under
  program Mobility Plus, grant no.1320/MOB/IV/2015/0.
\end{acknowledgements}

\bibliographystyle{spphys}
\bibliography{bibrefs}

\clearpage
\end{document}

%% file: figures.tex
\def\figNfPDF{
\begin{figure*}[tbh]
\centering{}
\includegraphics[width=0.45\textwidth]{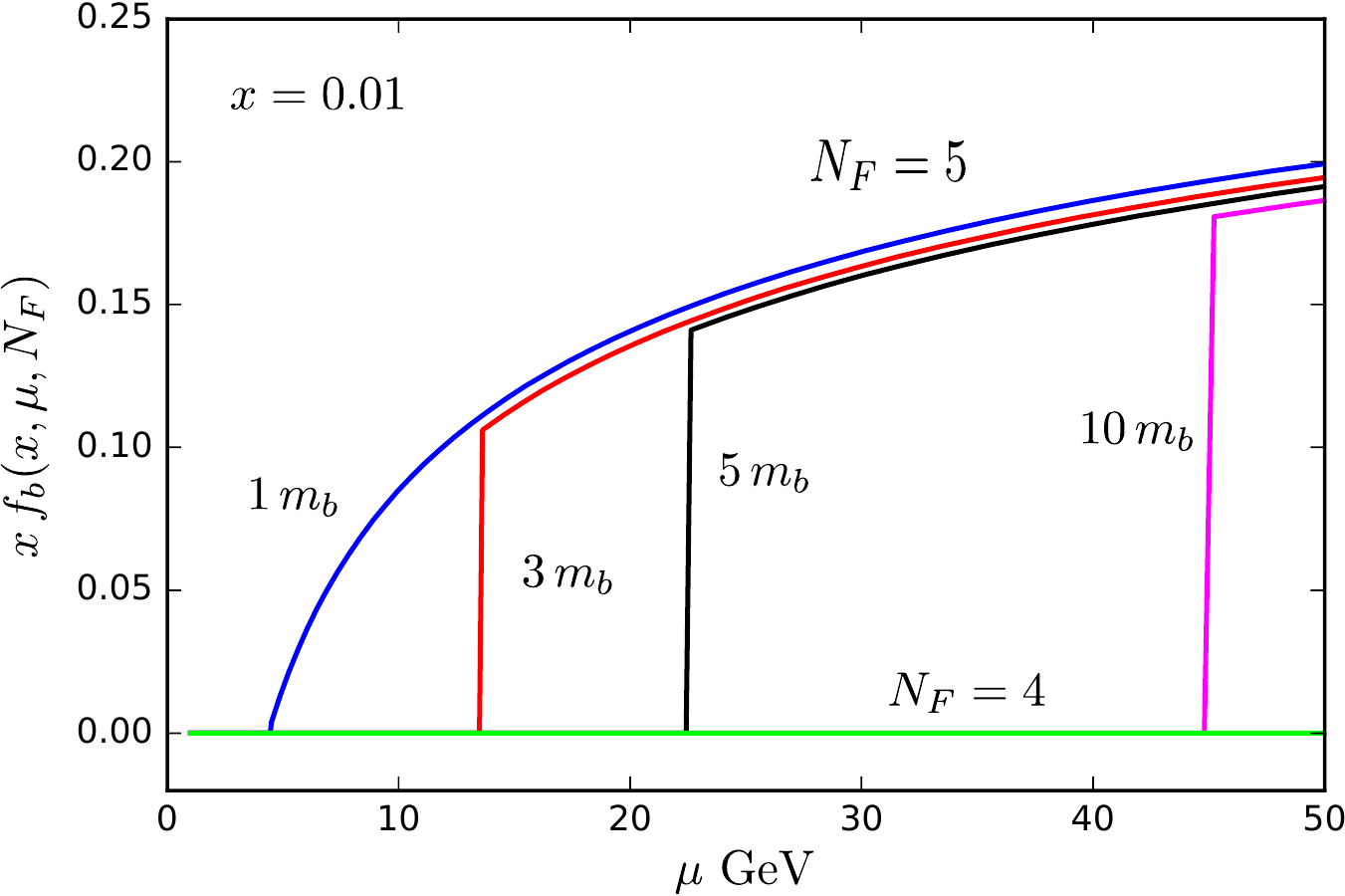}
\hfil
\includegraphics[width=0.45\textwidth]{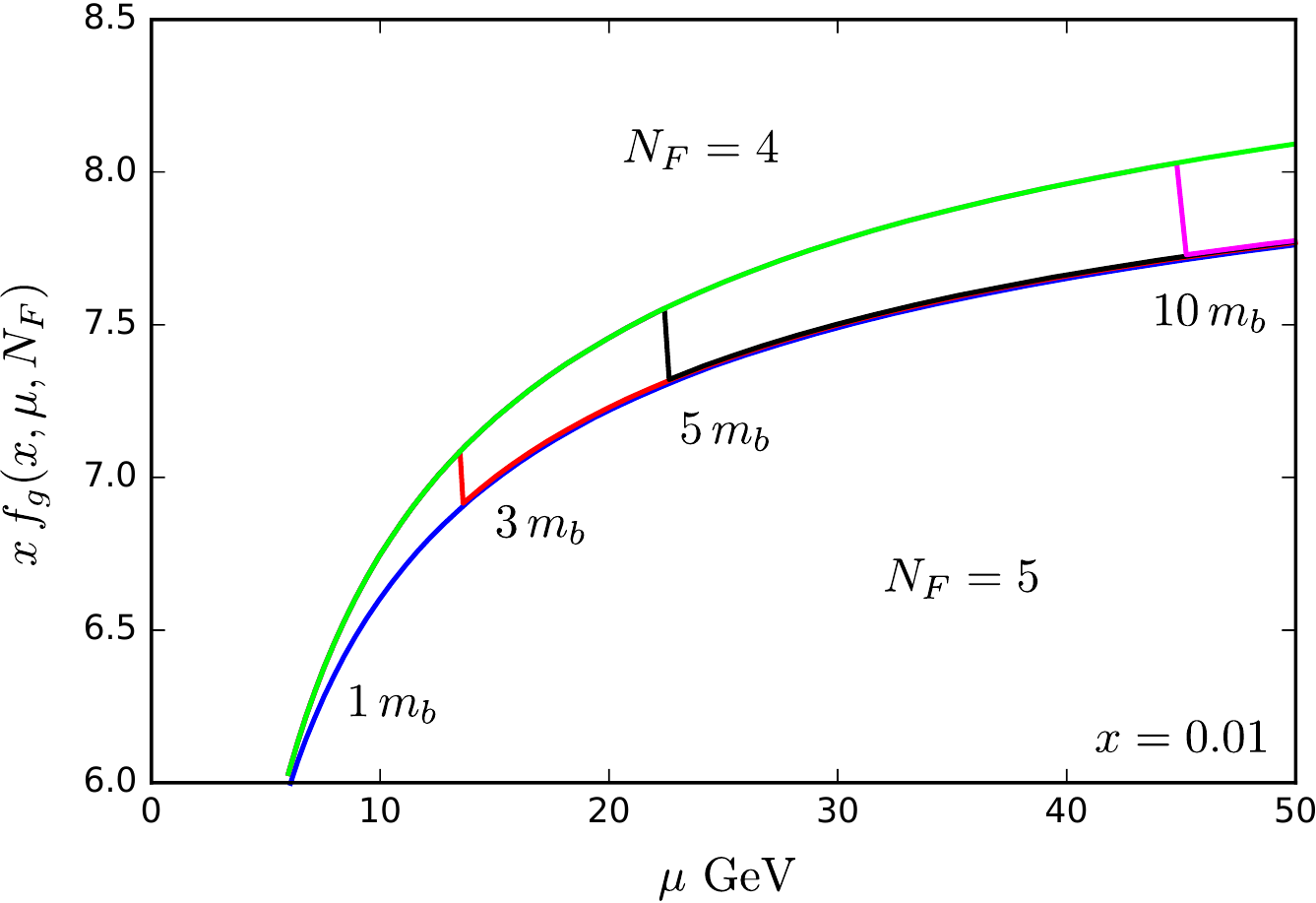}
\caption{$N_F$-dependent PDFs $x \, f_i(x,\mu,N_F)$ for the bottom quark (left) and gluon (right)
with  variable matching scales for 
$\mu_b=\{1,3,5,10, \infty\}\times m_b$ \{blue, red, black, magenta, green\}
with $x=0.01$ as a function of $\mu$ in GeV.  
The vertical lines in the plots show the transition from the $N_F=4$
to $N_F=5$ flavor scheme. 
\label{fig:pdfVsQ}
}
\end{figure*}
} %
\def\figvfnsi{
\begin{figure*}[tbh]
\centering{}
\includegraphics[width=0.45\textwidth]{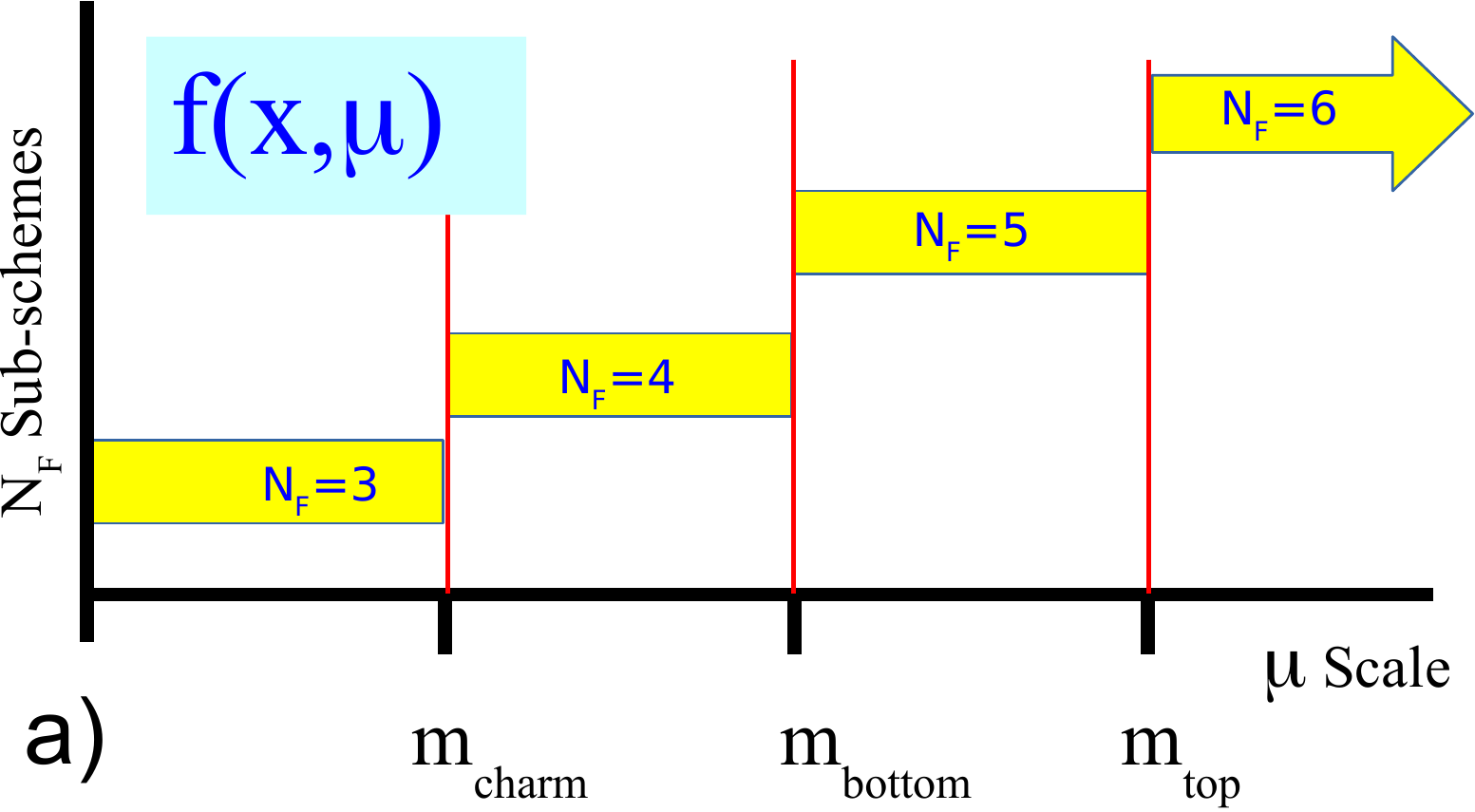}
\hfil
\includegraphics[width=0.45\textwidth]{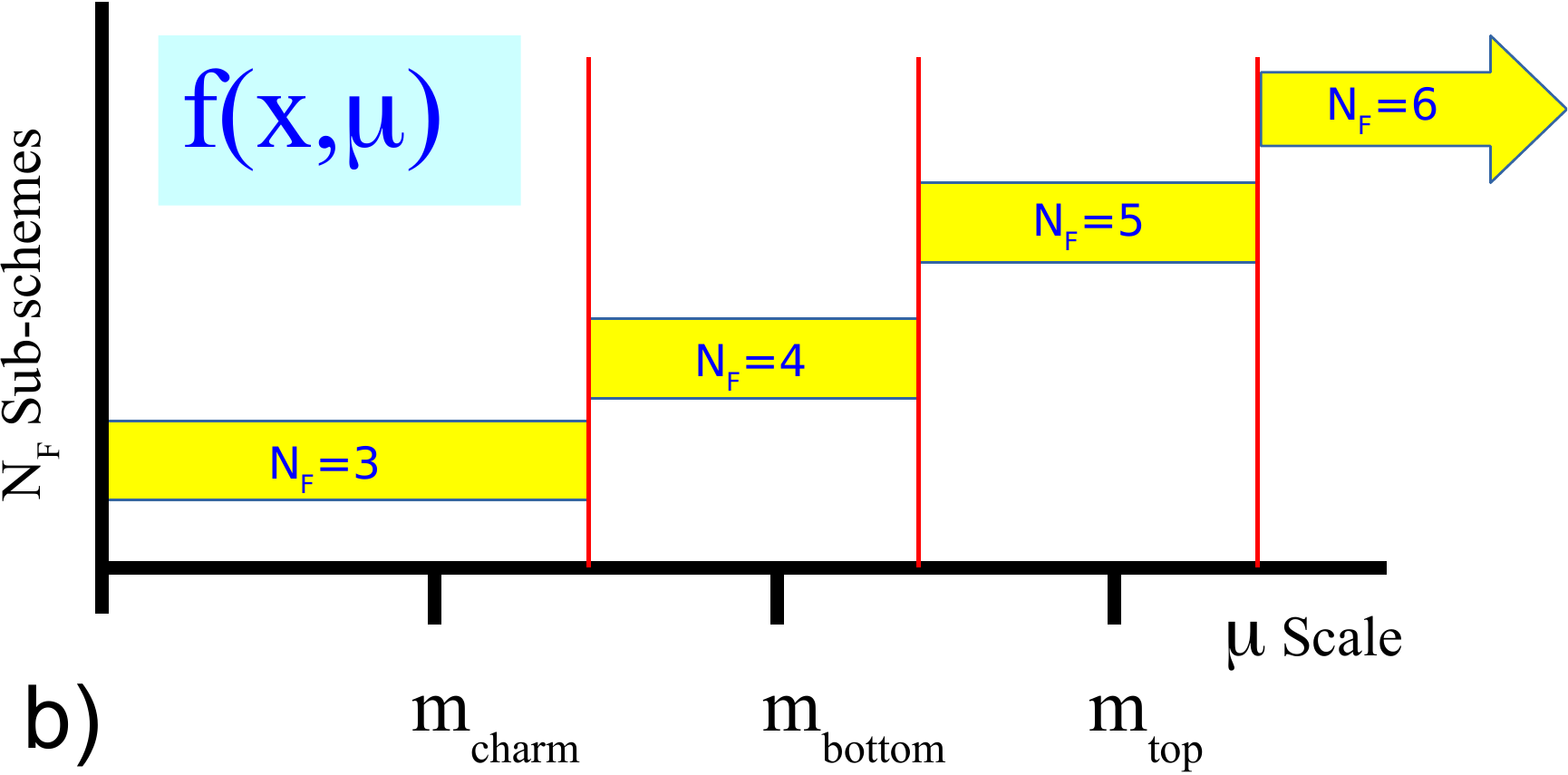}
\caption{An illustration of the separate $N_{F}$ renormalization sub-schemes
which define a VFNS. 
Historically, the matching scales $\mu_{m}$ 
were chosen to be exactly the mass values $m_{c,b,t}$ as in Figure-a.
\quad 
Figure-b is a generalized case where the matching scales $\mu_{m}$
are chosen to be different from the mass values. \label{fig:vfns1}
}
\end{figure*}
} %
\def\figacot{
\begin{figure*}[tbh]
\centering{}
\includegraphics[width=0.95\textwidth]{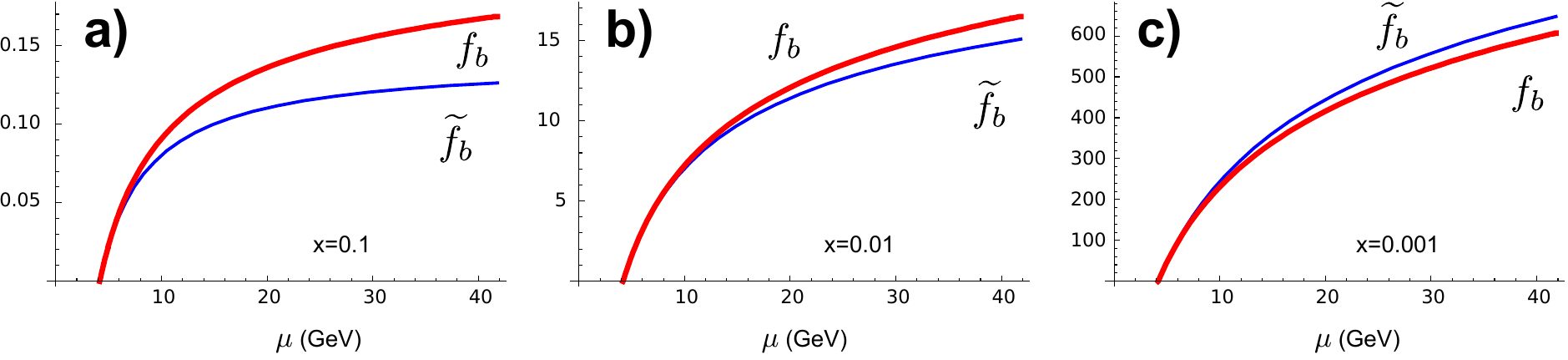}
\caption{The comparison of the DGLAP evolved PDF $f_b(x,\mu)$ and 
the perturbatively calculated  $\widetilde{f}_b(x,\mu)$
as a function of $\mu$ for selected $x$ values.
For $\mu\to m_b$ we find the functions match precisely: 
$\widetilde{f}_b(x,\mu) \to f_b(x,\mu)$.
We have used \hbox{NNPDF30\_lo\_as\_118\_nf\_6} as the base  PDF set.
 \label{fig:acot}
}
\end{figure*}
} %
\def\figBmatch{
\begin{figure*}[tbh]
\centering{}
\includegraphics[width=0.45\textwidth,angle=0]{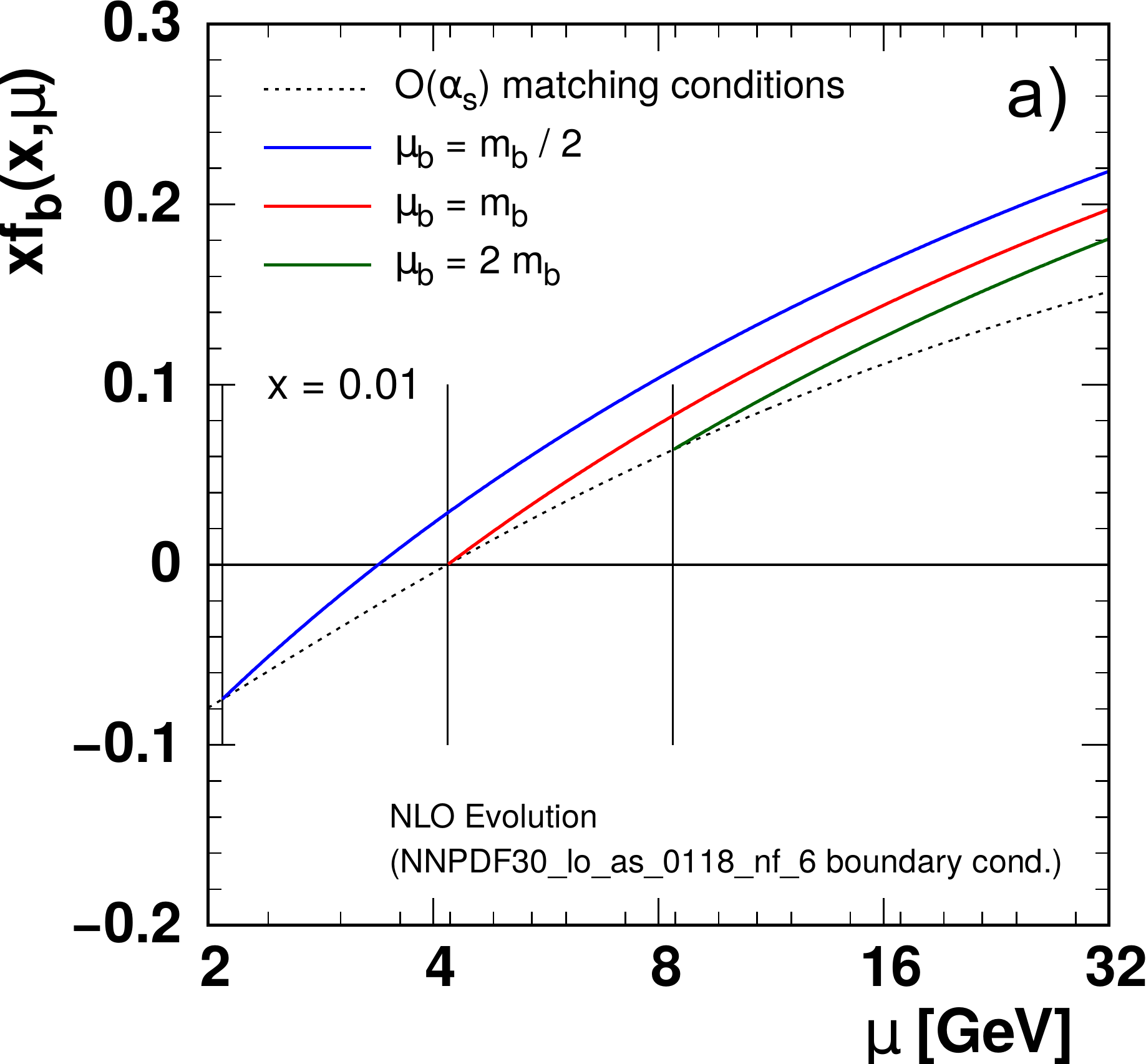}
\hfil
\includegraphics[width=0.45\textwidth,angle=0]{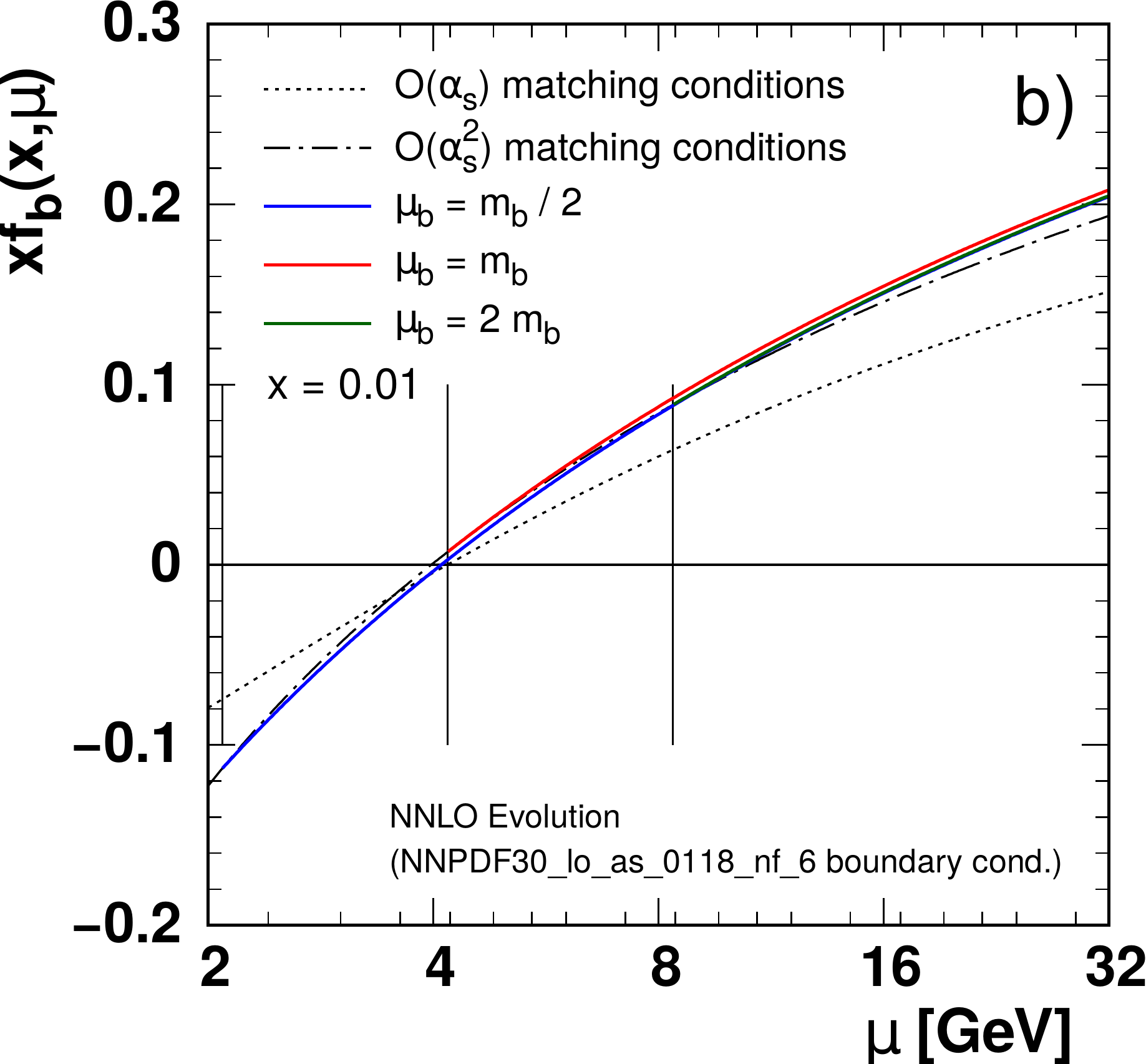}
\caption{We display the b-quark PDF $x\, f_b^{(5)}(x,\mu)$ for different choices
  of the matching scales 
$\mu_{m}=\{m_b/2,m_b,2m_b \}$ (indicated by the vertical lines)
computed at 
NLO (Fig.-a)
and NNLO (Fig.-b).
\label{fig:bMatch}
}
\end{figure*}
} %
\def\figFbottom{
\begin{figure*}[tbh]
\centering{}
\includegraphics[width=0.45\textwidth]{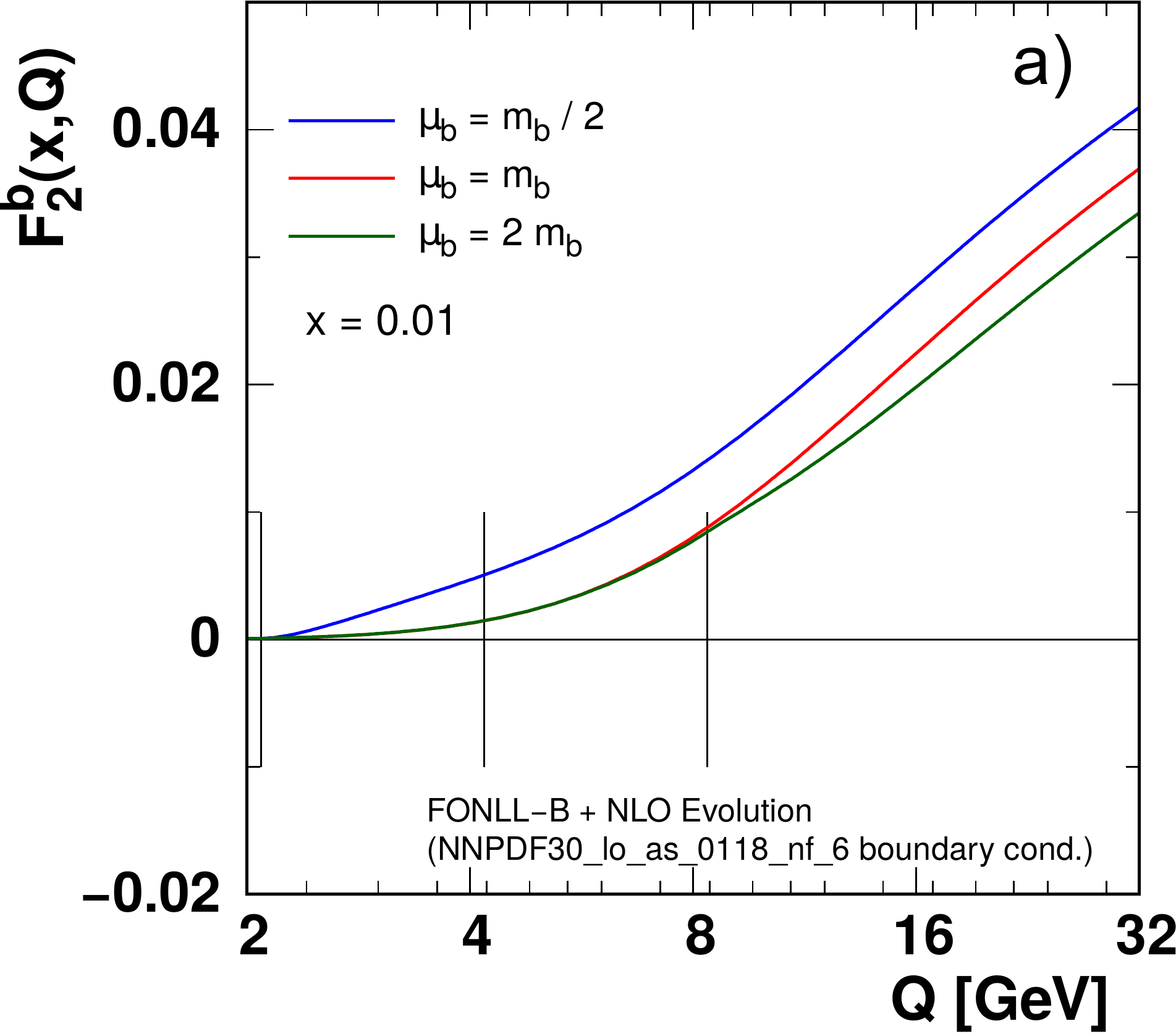}
\hfil
\includegraphics[width=0.45\textwidth]{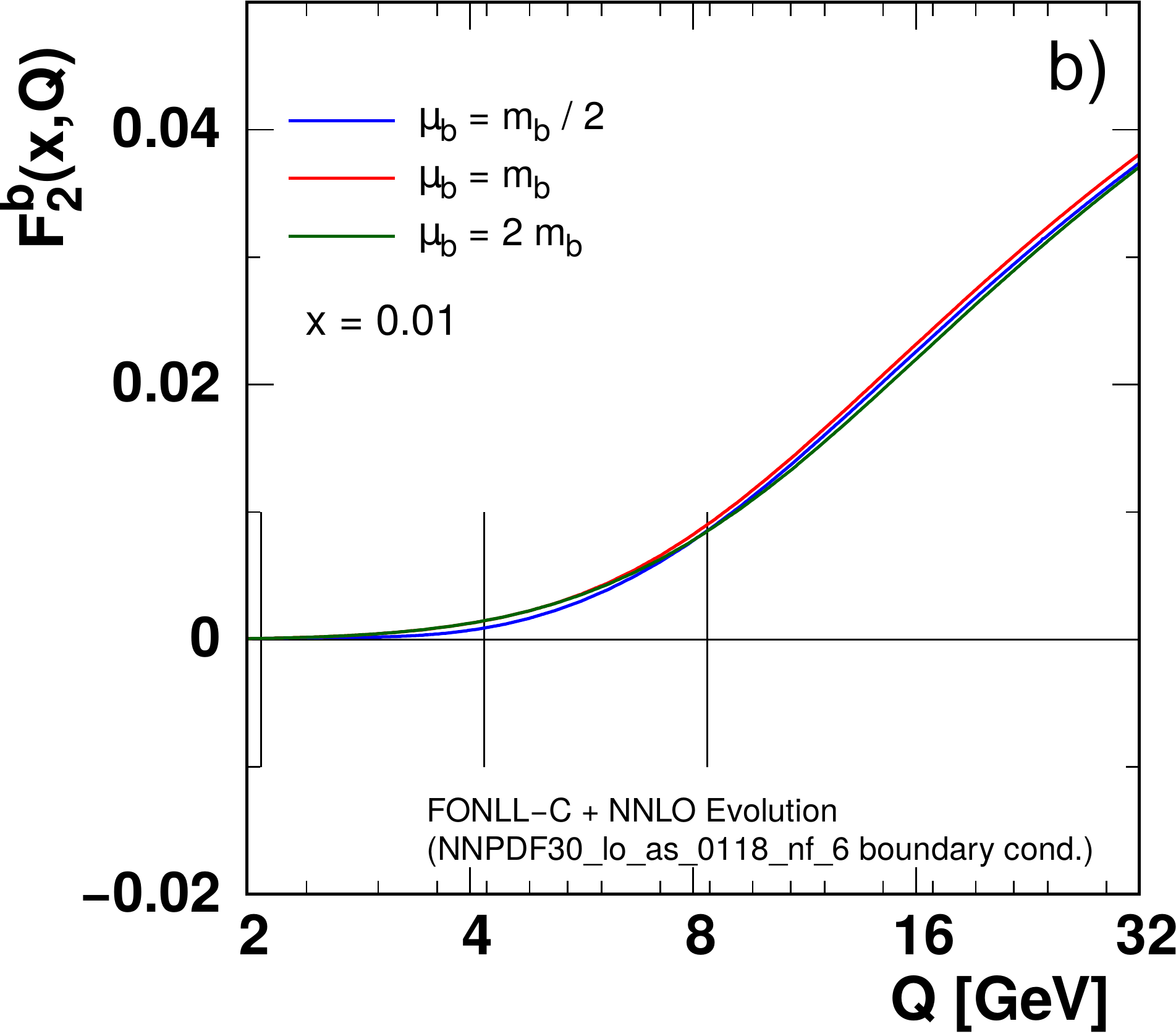}
\caption{We display$F_2^b(x,Q)$  for different choices
  of the matching scales 
$\mu_{m}=\{m_b/2,m_b,2m_b \}$ (indicated by the vertical lines)
computed at 
NLO (Fig.-a)
and NNLO (Fig.-b).
Here, we have chosen $\mu=Q$.
For details on the FONNL calculation see Ref.~\cite{Forte:2010ta}.
\label{fig:fbottom}
}
\end{figure*}
} %
\def\figCharm{
\begin{figure*}[tbh]
\centering{}
\includegraphics[width=0.42\textwidth]{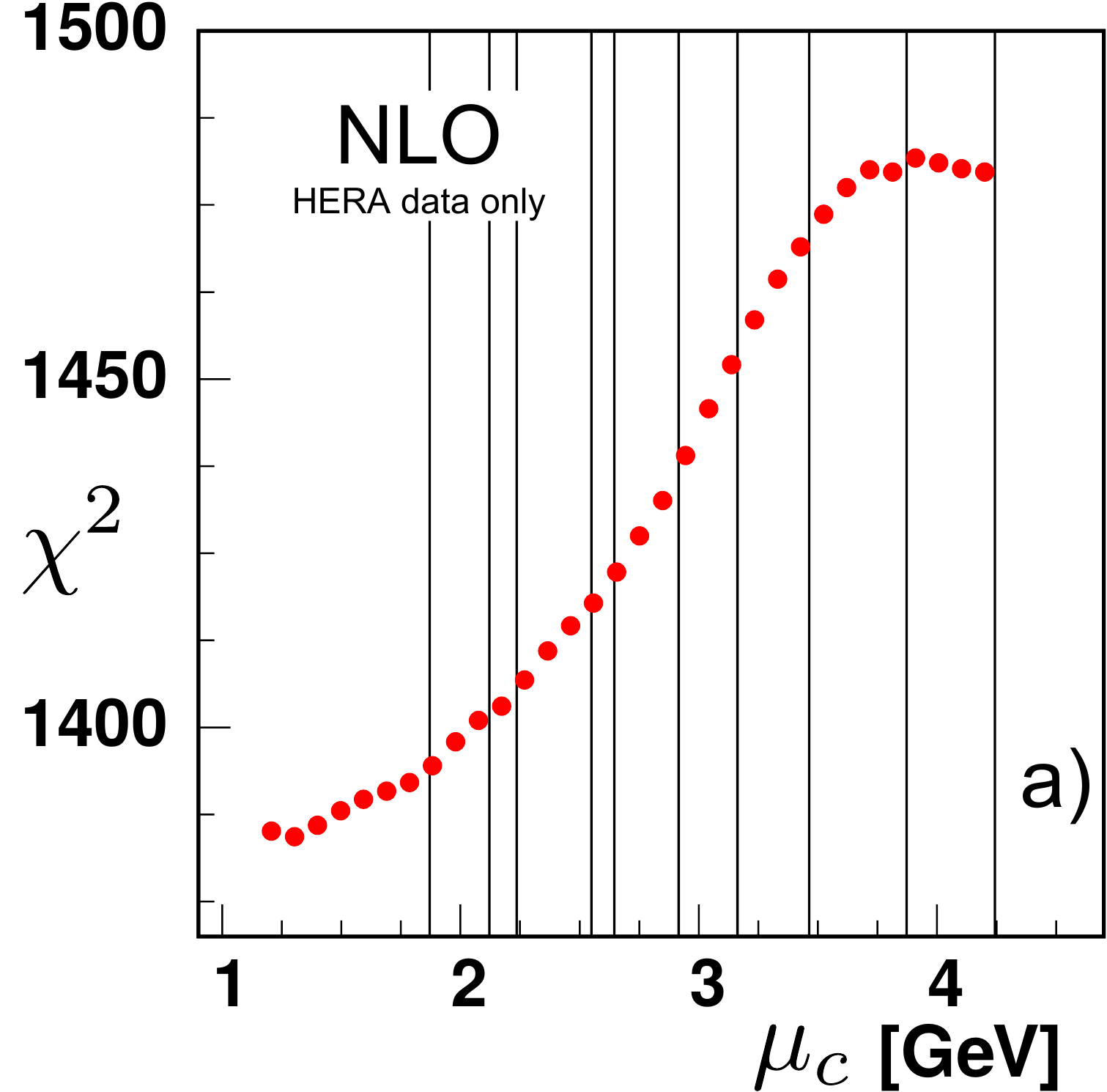}
\hfil
\includegraphics[width=0.45\textwidth]{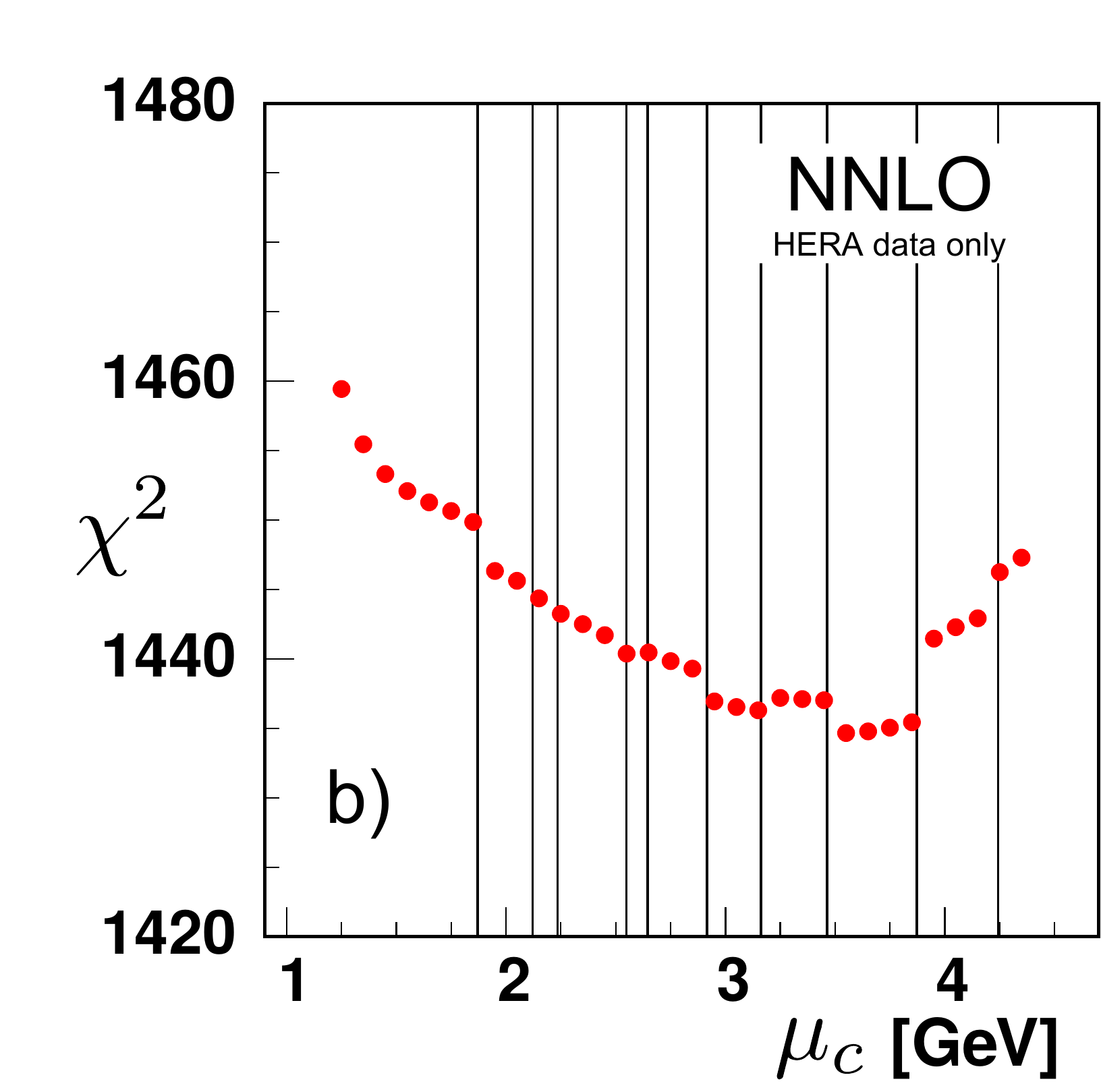}
\caption{$\chi^{2}$ vs. the charm matching scale $\mu_{c}$
at a) NLO and b) NNLO  for all data sets. 
The bin boundaries for the HERA data set \hbox{``HERA1+2 NCep 920''}
are indicated by the vertical lines.
\label{fig:charm}}
\end{figure*}
} %
\def\figCharmOnly{
\begin{figure*}[tbh]
\centering{}
\includegraphics[width=0.45\textwidth]{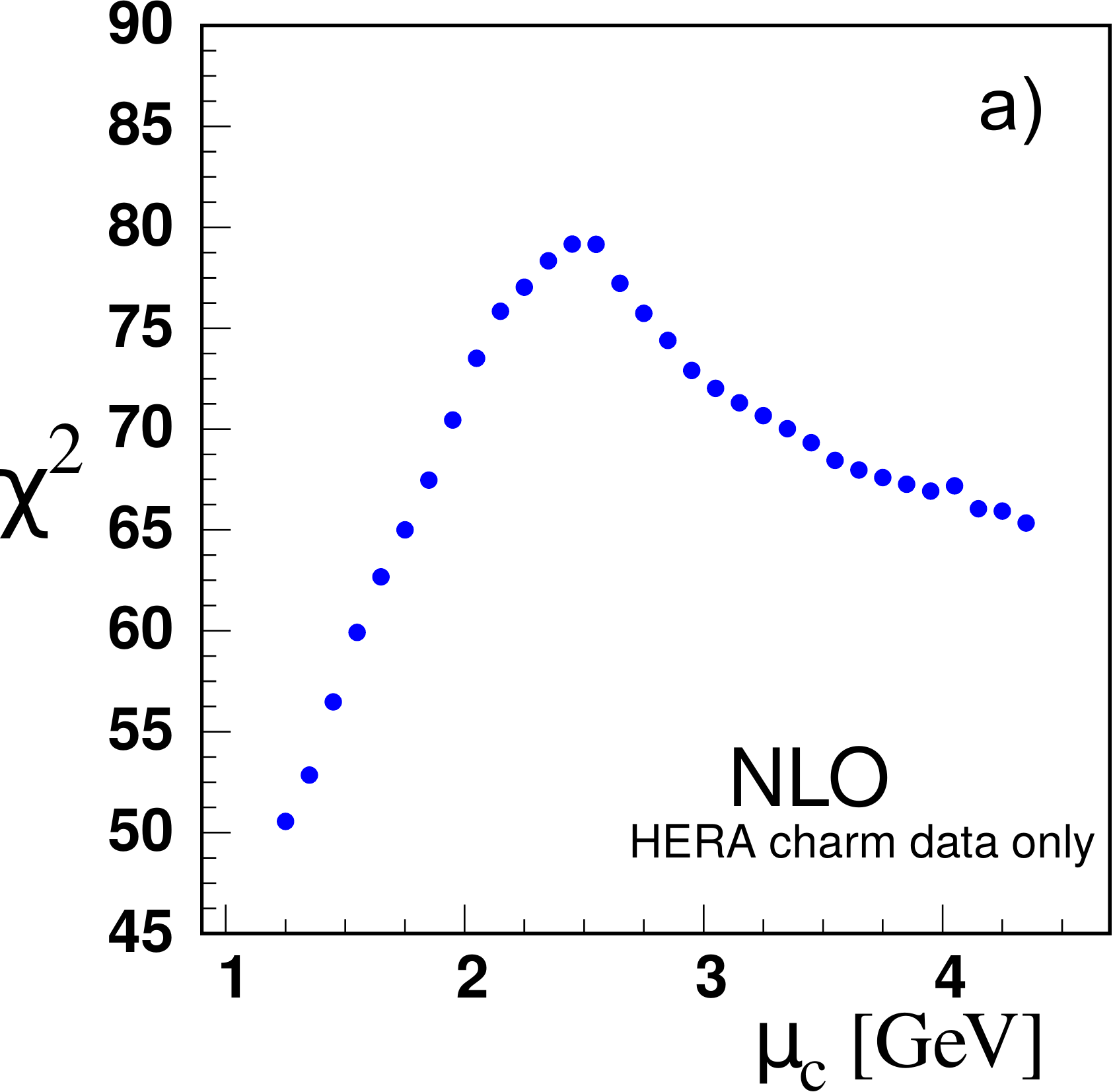}
\hfil
\includegraphics[width=0.45\textwidth]{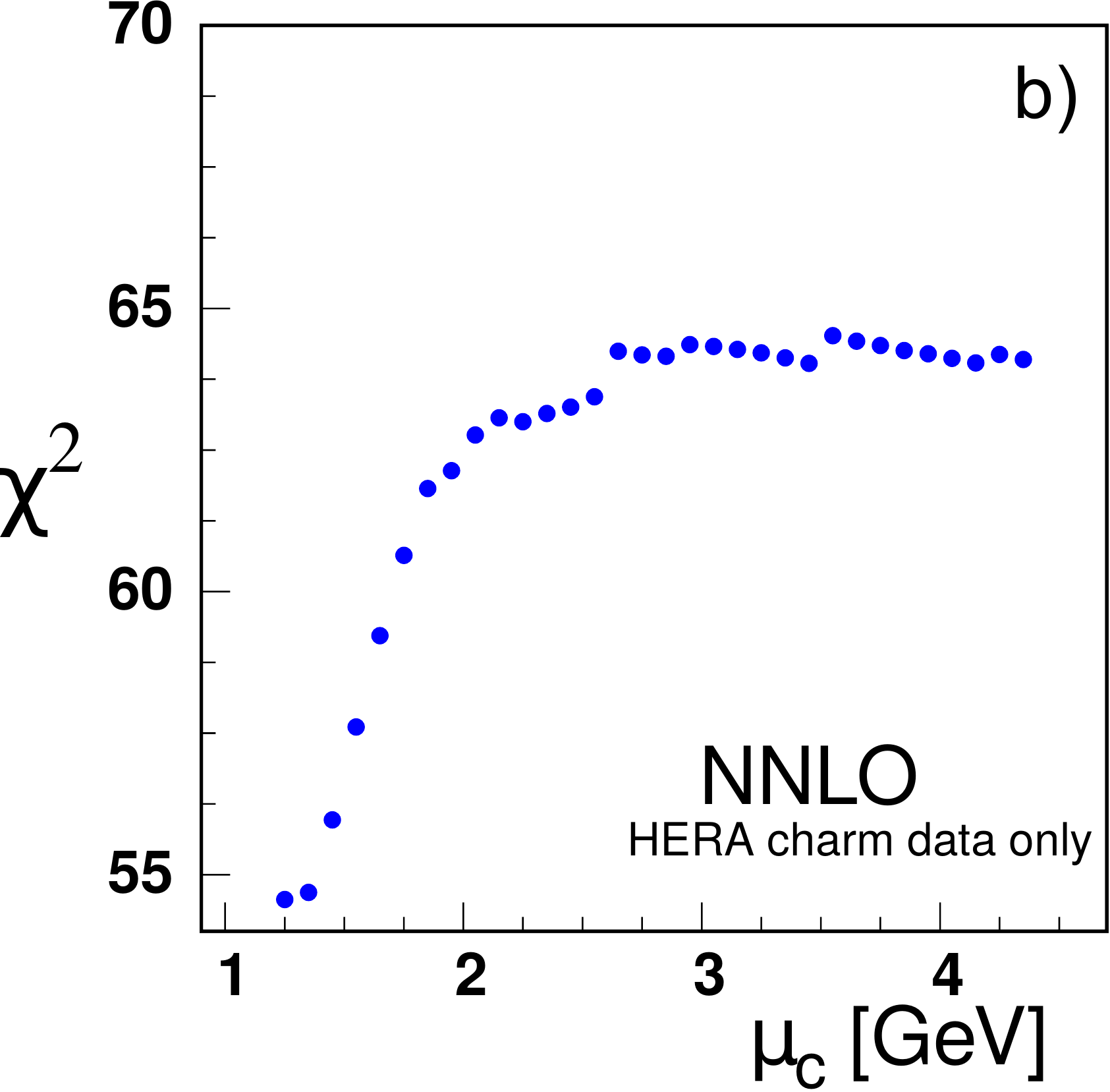}
\caption{$\chi^{2}$ vs. the charm matching scale $\mu_{c}$
at a) NLO and b) NNLO for only the H1-ZEUS combined charm production data;
note, this includes the correlated $\chi^2$ contribution from Tables~\ref{tab:charmNLO}
and~\ref{tab:charmNNLO}.
\label{fig:charmOnly}}
\end{figure*}
} %
\def\figBottom{
\begin{figure*}[tbh]
\centering{}
\includegraphics[width=0.45\textwidth]{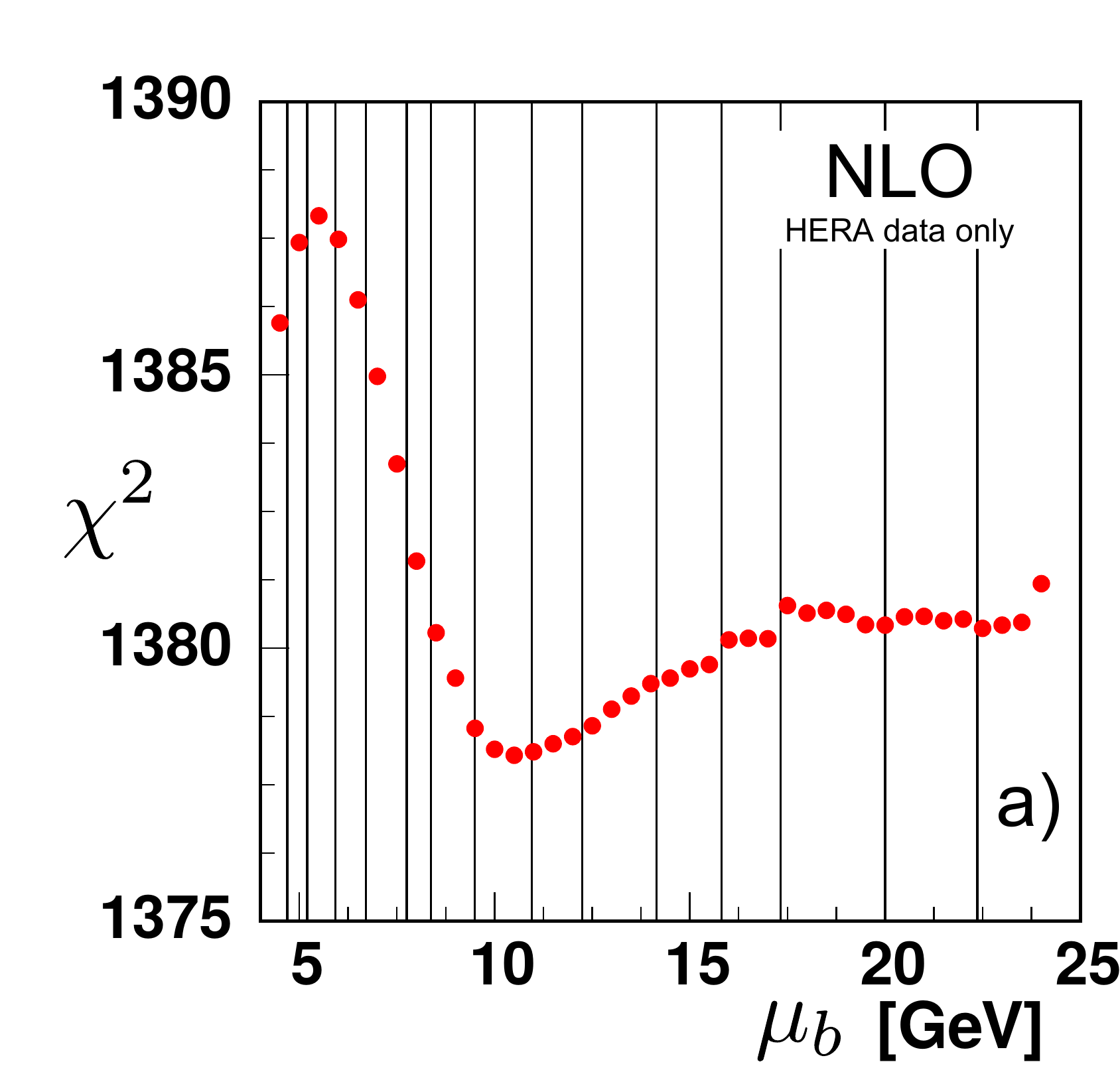}
\hfil
\includegraphics[width=0.45\textwidth]{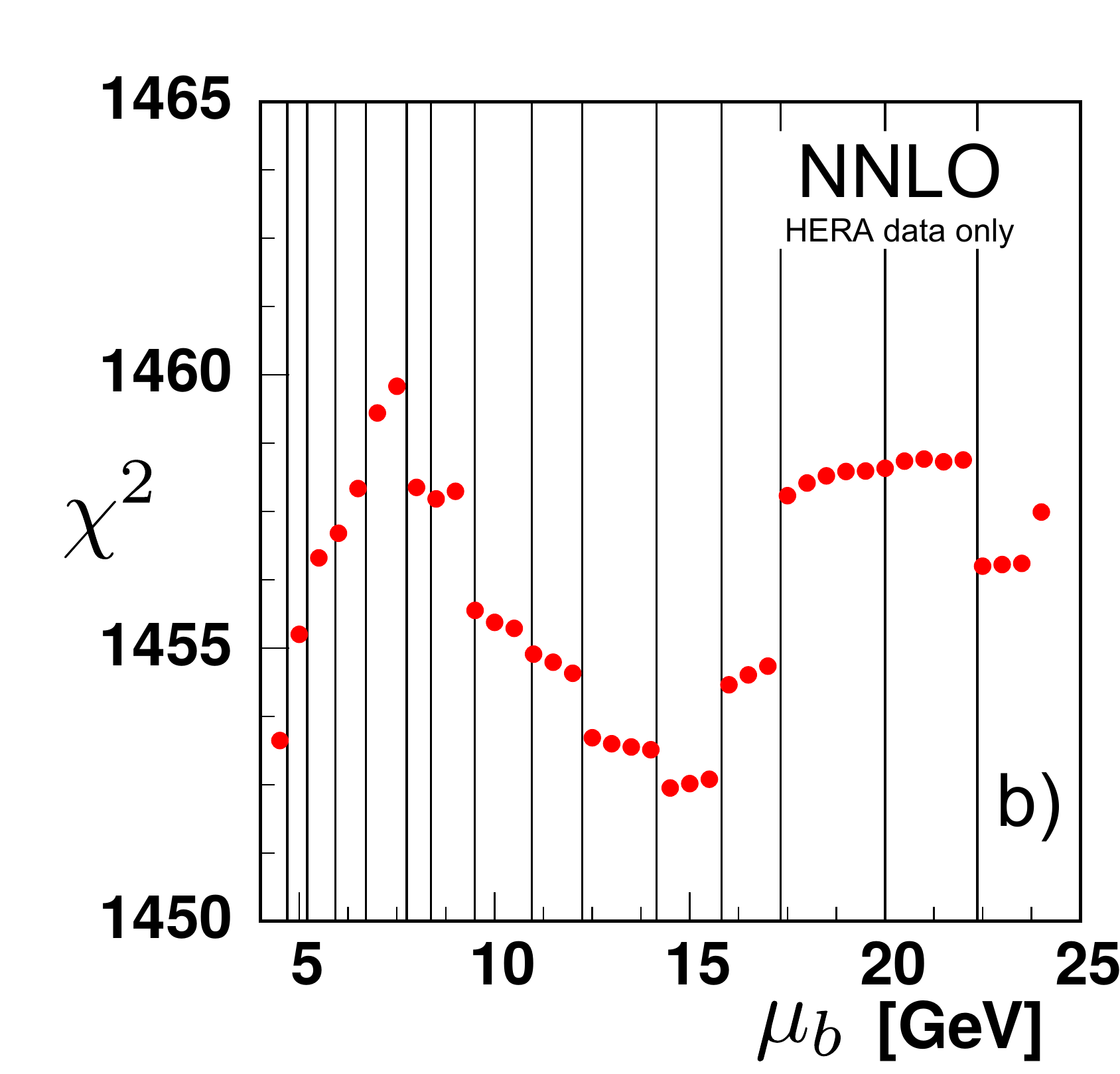}
\caption{$\chi^{2}$  vs. the bottom matching scale $\mu_{b}$
at a) NLO and b) NNLO  for all data sets. 
The bin boundaries for the HERA data set \hbox{``HERA1+2 NCep 920''}
are indicated by the vertical lines.
\label{fig:bottom}}
\end{figure*}
} %
\def\figBottomOnly{
\begin{figure*}[tbh]
\centering{}
\includegraphics[width=0.42\textwidth]{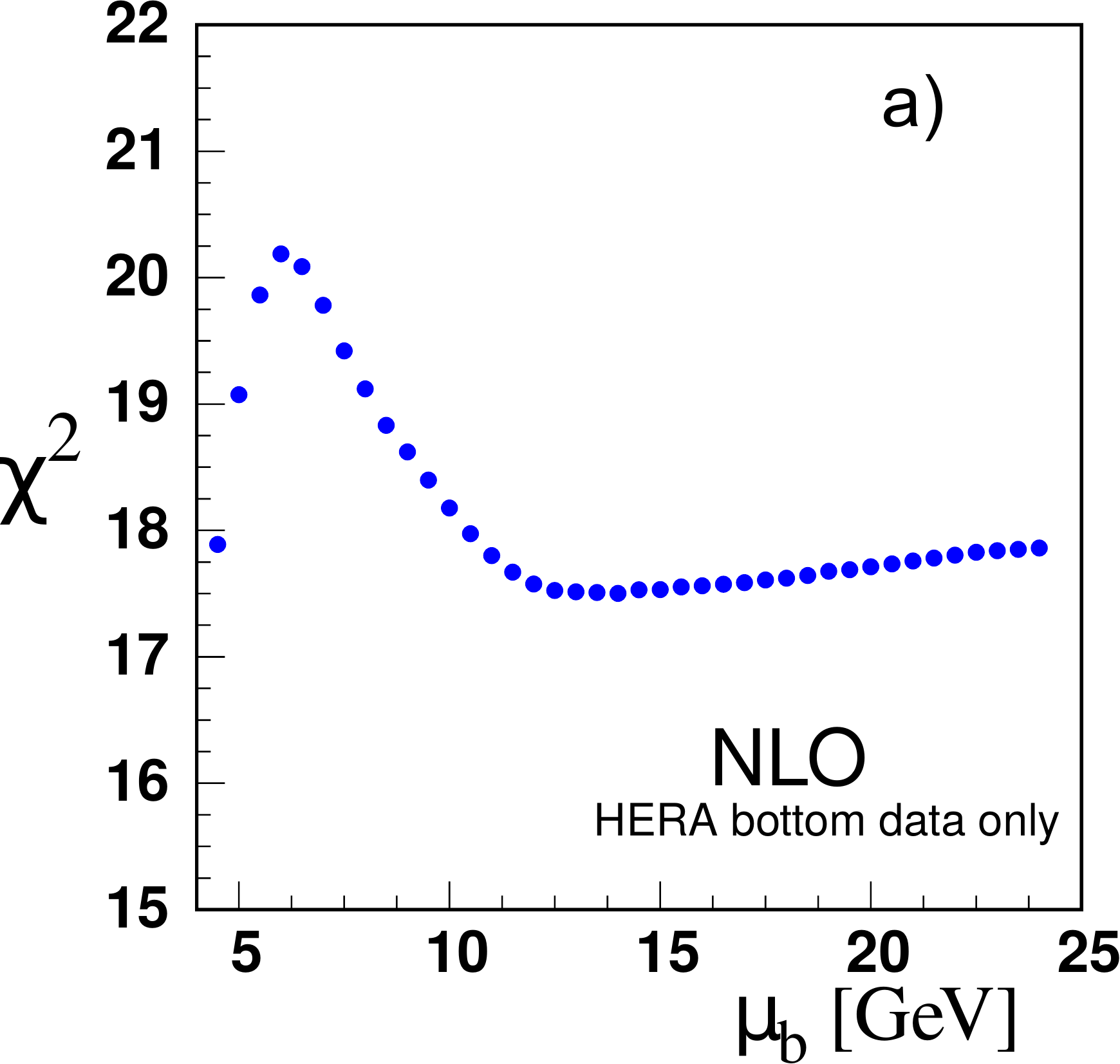}
\hfil
\includegraphics[width=0.42\textwidth]{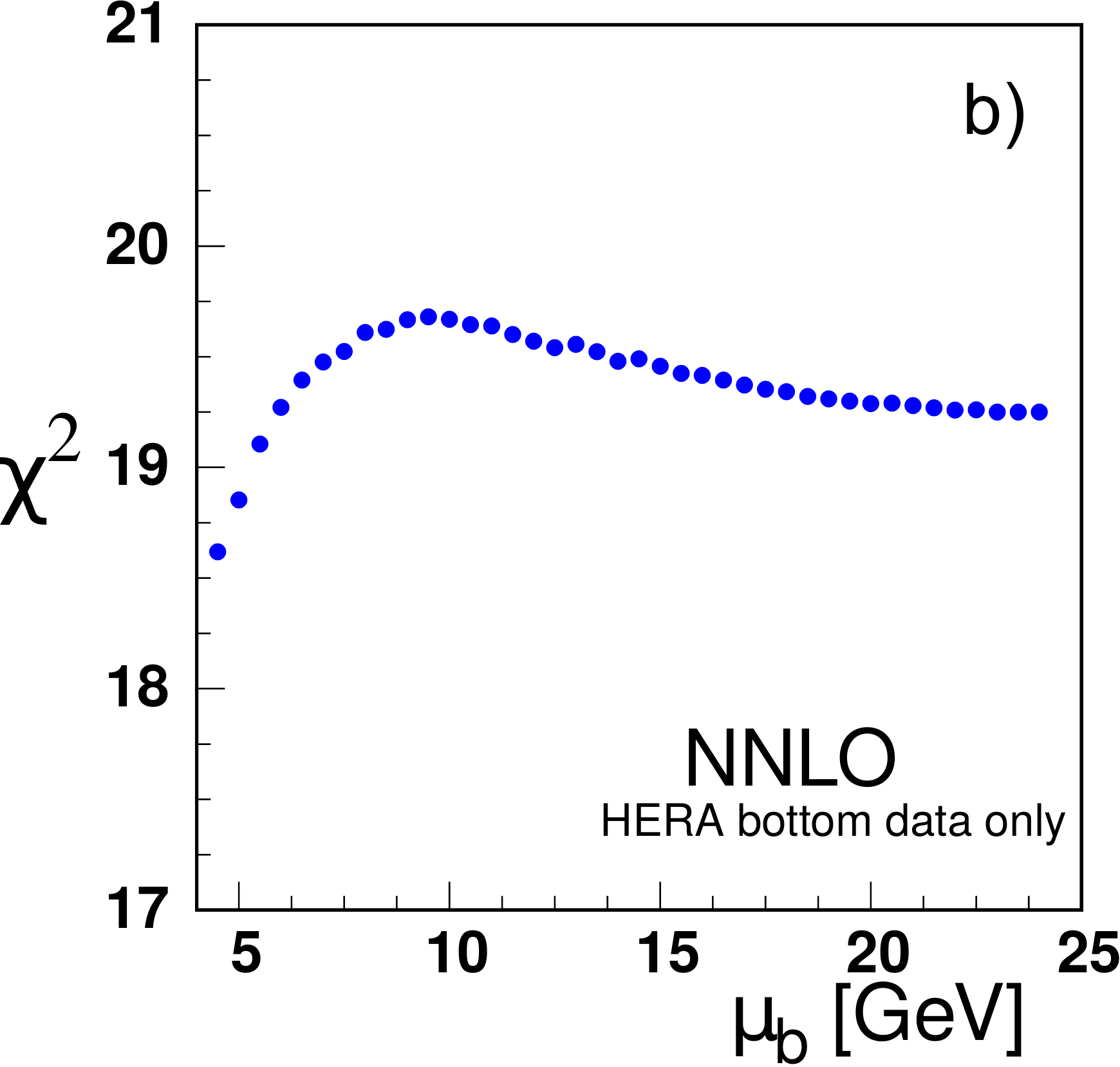}
\caption{
$\chi^{2}$ vs.  the bottom matching scale $\mu_{b}$
at a) NLO and b) NNLO for only the bottom data;
note, this includes the H1 and ZEUS beauty data as well
as the correlated $\chi^2$ contribution from Tables~\ref{tab:bottomNLO}
and~\ref{tab:bottomNNLO}.
\label{fig:bottomOnly}}
\end{figure*}
} %
\def\figChiScaledi{
\begin{figure*}[tbh]
\centering{}
\includegraphics[width=0.45\textwidth]{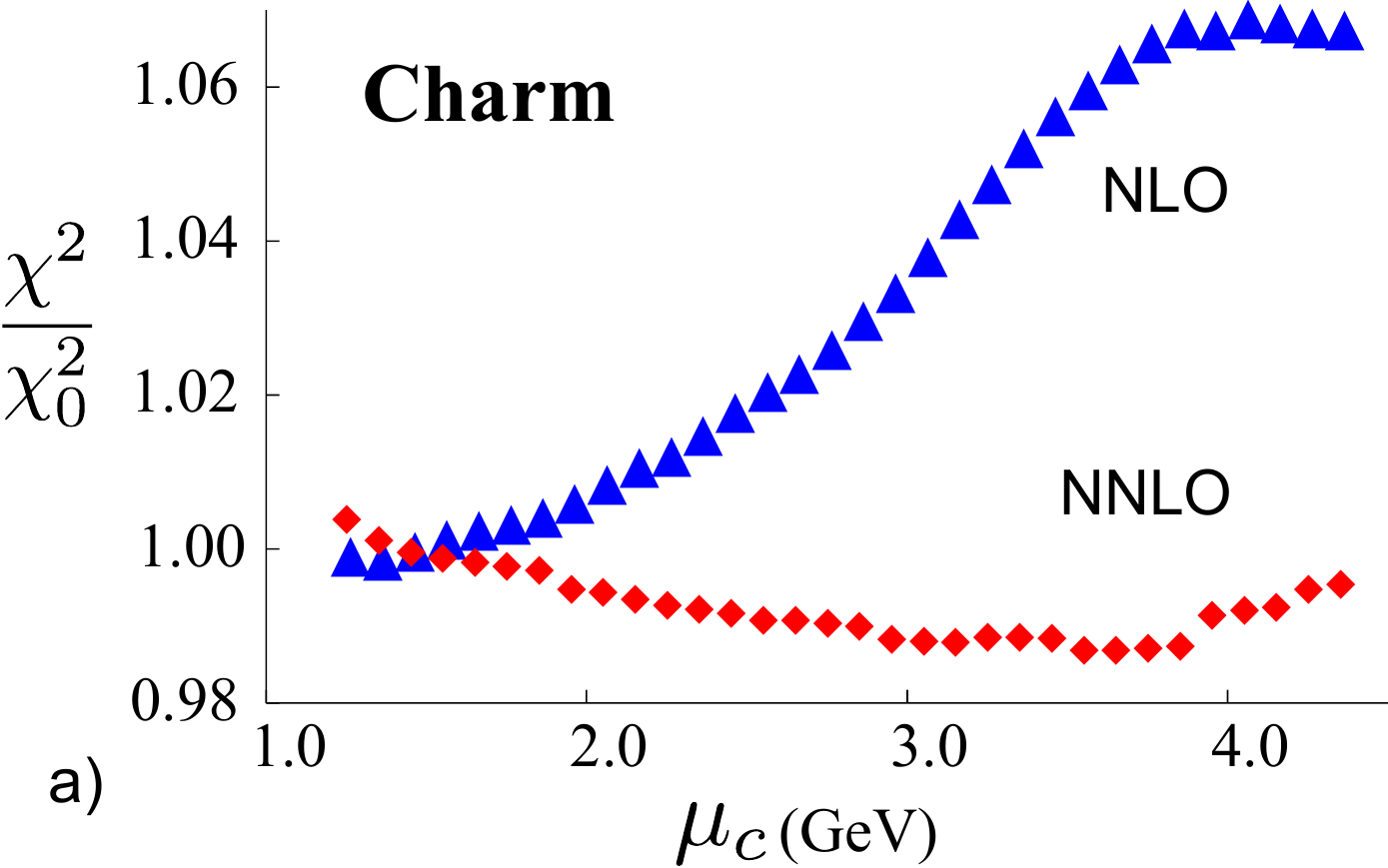}
\hfil
\includegraphics[width=0.45\textwidth]{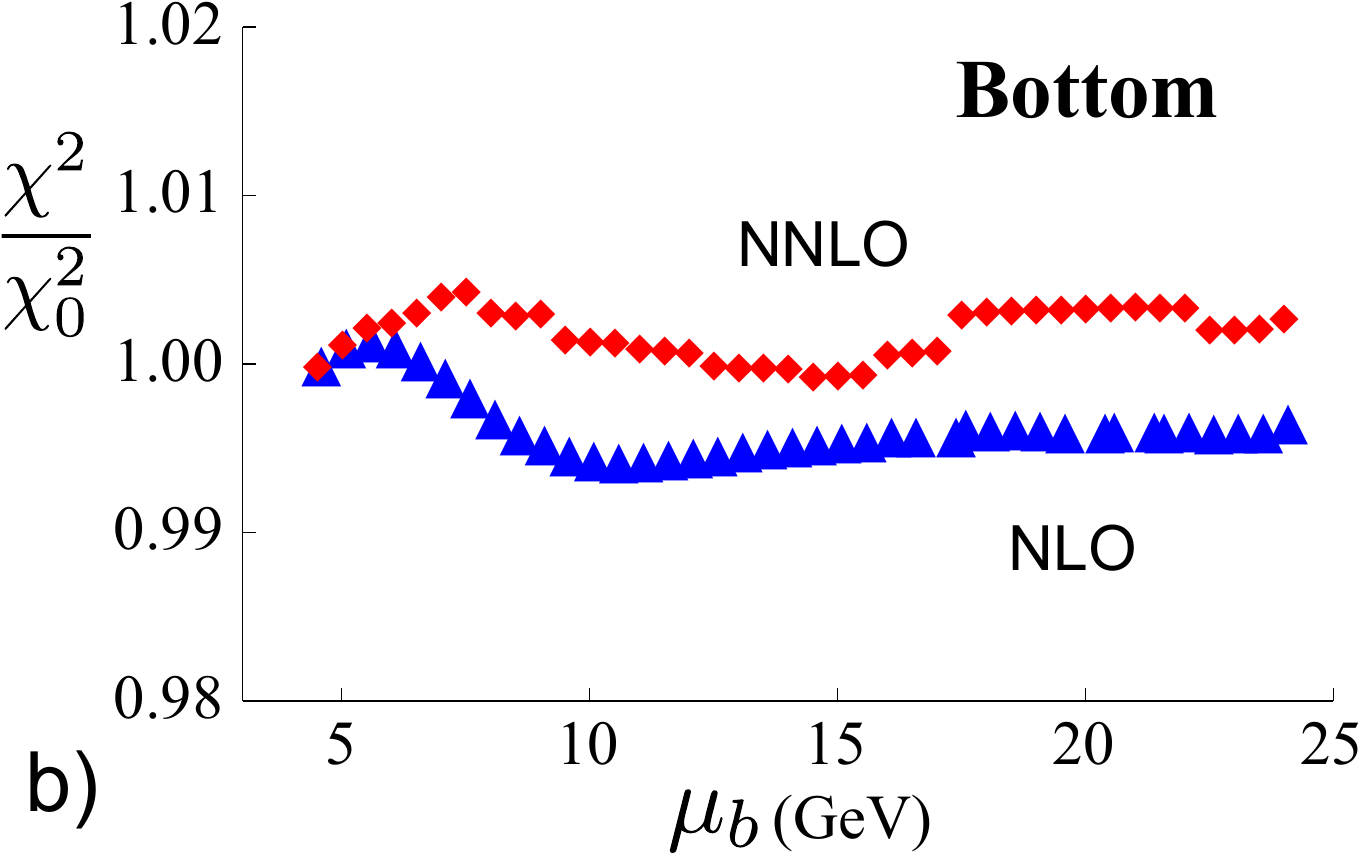}
\caption{
The ratio ($\chi^2/\chi^2_0$) of total $\chi^2$ values
(all data sets combined) from Figs.~\ref{fig:charm} and~\ref{fig:bottom},
as a function of the a) charm and b) bottom  matching scale $\mu_{c,b}$ in GeV. 
$\chi_{0}^{2}$ is the $\chi^{2}$ value for $\mu_{m}$ equal to the quark mass. 
The triangles (blue  $\blacktriangle$ )  are NLO
and the diamonds (red $\blacklozenge$) are NNLO. 
\label{fig:chi2scaledi}
}
\end{figure*}
} %
\def\figChiScaledii{
\begin{figure*}[tbh]
\centering{}
\includegraphics[width=0.45\textwidth]{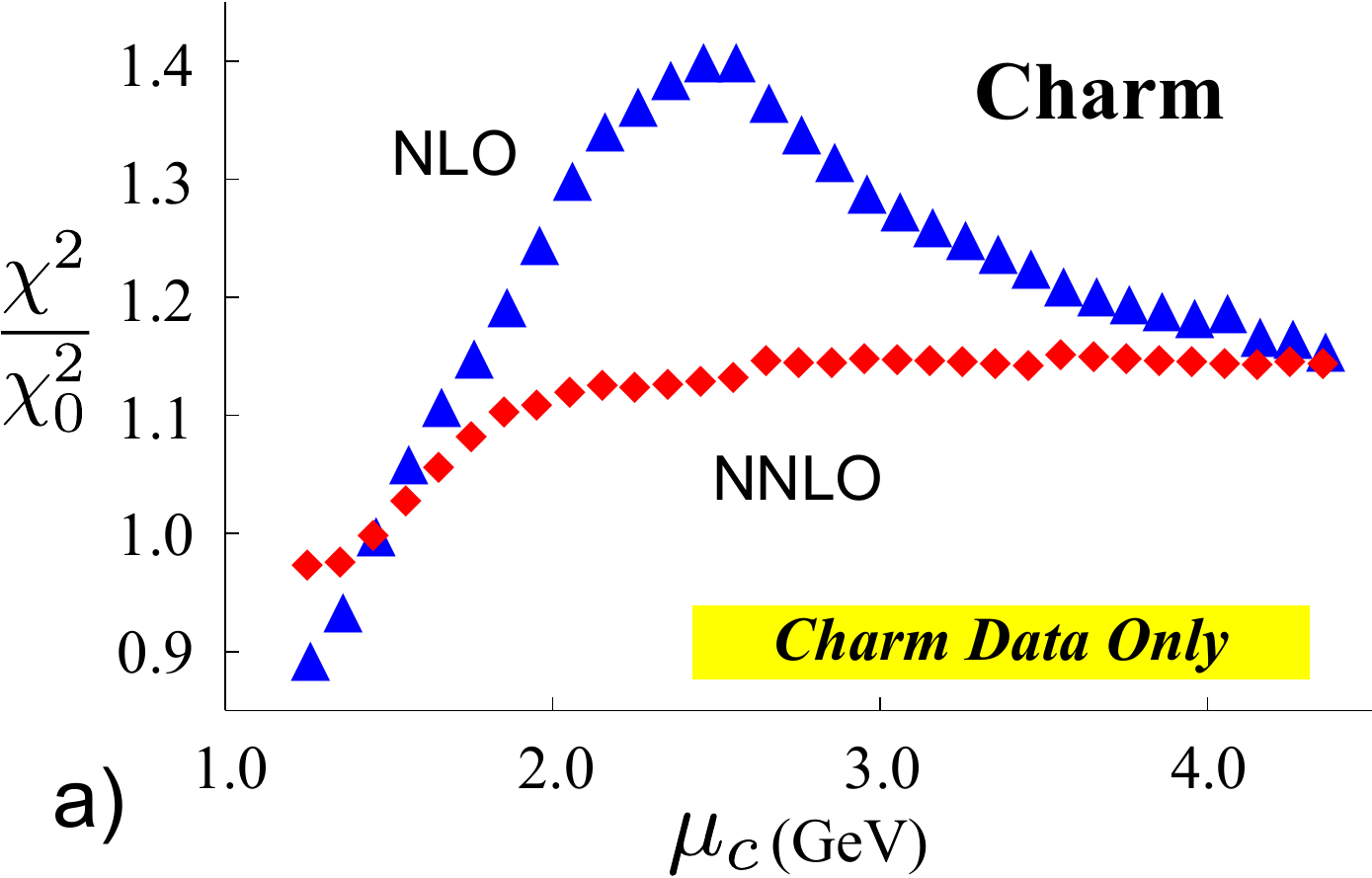}
\hfil
\includegraphics[width=0.45\textwidth]{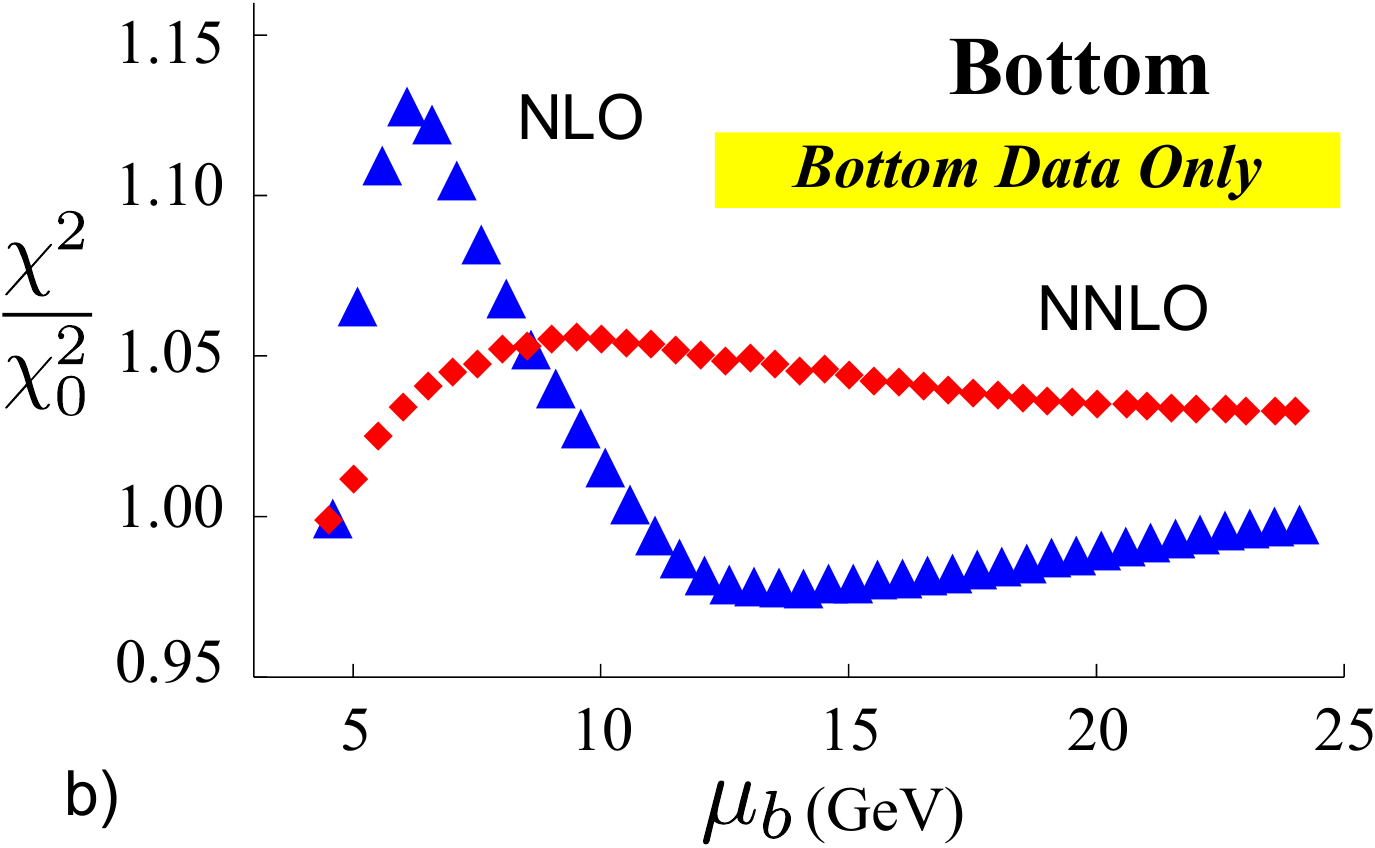}
\caption{
The ratio ($\chi^2/\chi^2_0$) of partial $\chi^2$ values 
(charm/bottom data only) from Figs.~\ref{fig:charmOnly} and~\ref{fig:bottomOnly} 
as a function of the a) charm and b) bottom  matching scale $\mu_{c,b}$ in GeV. 
$\chi_{0}^{2}$ is the $\chi^{2}$ value for $\mu_{m}$ equal to the quark mass.
The triangles (blue  $\blacktriangle$ )  are NLO
and the diamonds (red $\blacklozenge$) are NNLO. 
\label{fig:chi2scaledii}
}
\end{figure*}
} %
\def\figvfnsii{
\begin{figure}[tbh]
\centering{}
\includegraphics[width=0.45\textwidth]{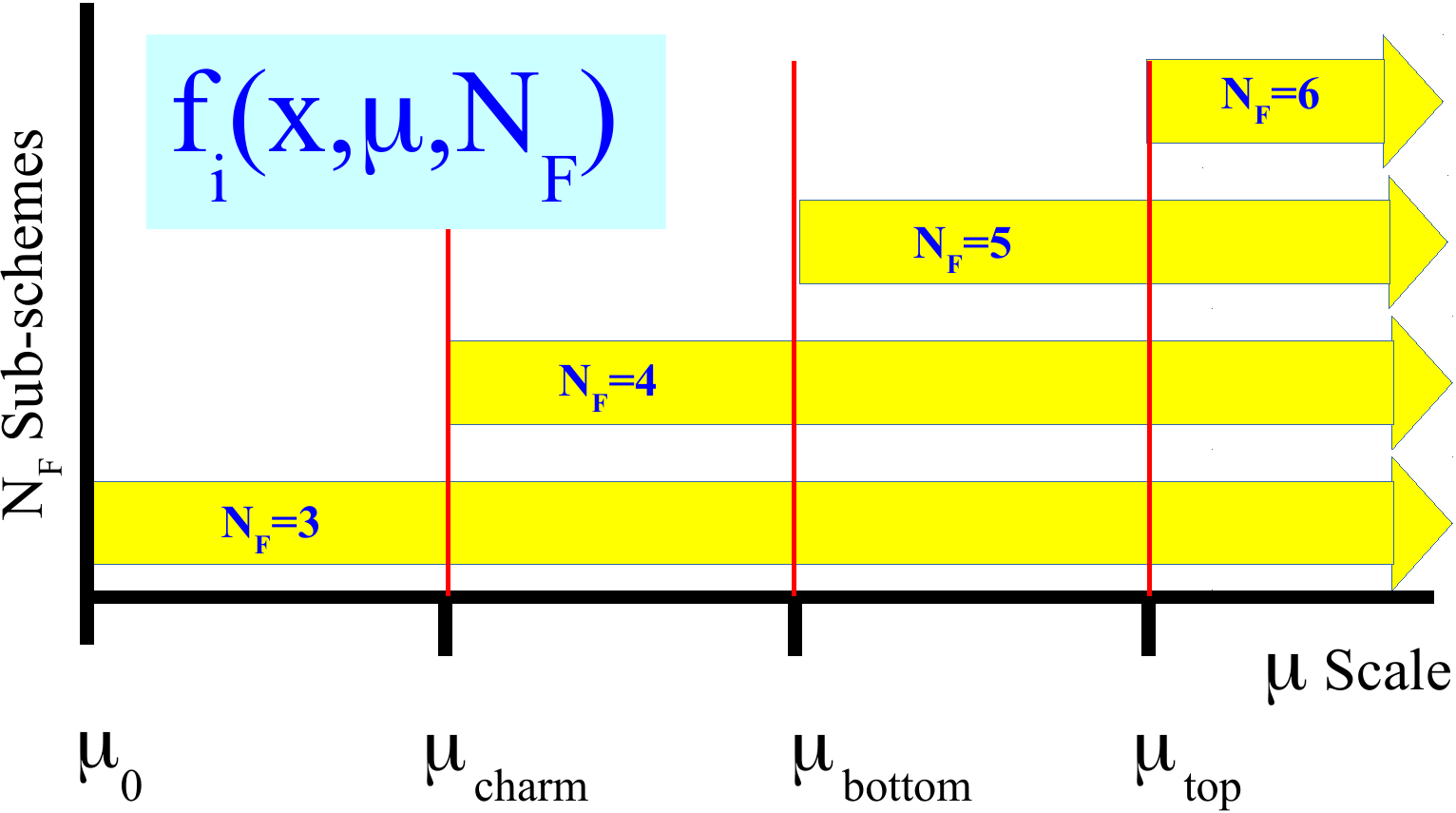}
\caption{An illustration of the separate $N_{F}$ renormalization sub-schemes
which define the VFNS. 
In contrast to Fig.~\ref{fig:vfns1}-a), each of the $N_{F}$ sub-schemes are available
for all scales above $\mu_{m}$. 
The particular scheme can be specified
by choosing $N_{F}$ when calling the PDF, i.e. $f_{i}(x,\mu,N_{F})$.
This illustration shows a matching scale  of $\mu_{m}=m_H$.
\label{fig:vfns2}}
\end{figure}
} %

%% file: tables.tex
\def\tableCharmNLO{             
\begin{table*}[t]             
   \begin{center}
  \rowcolors{2}{lightgray}{}
    \begin{tabular}{lp{1.65cm}p{1.65cm}p{1.65cm}p{1.65cm}p{1.65cm}}
      \toprule
       {\bf Charm NLO}   & \hspace{-0.35in} $\mu_c = \quad  1\, m_c$   & $2\, m_c$   & $3\, m_c$       \\ 
      \midrule
  Charm cross section H1-ZEUS combined \cite{Abramowicz:1900rp} & 46 / 47& 61 / 47& 54 / 47  \\ 
  H1 F2 Beauty Vertex \cite{Aaron:2009af} & 3.1 / 12& 2.8 / 12& 2.7 / 12  \\ 
  Beauty cross section ZEUS Vertex \cite{Abramowicz:2014zub} & 12 / 17& 12 / 17& 12 / 17  \\ 
  HERA1+2 CCep \cite{Abramowicz:2015mha} & 44 / 39& 44 / 39& 45 / 39  \\ 
  HERA1+2 CCem \cite{Abramowicz:2015mha} & 52 / 42& 47 / 42& 48 / 42  \\ 
  HERA1+2 NCem \cite{Abramowicz:2015mha} & 220 / 159& 228 / 159& 227 / 159  \\ 
  HERA1+2 NCep 820 \cite{Abramowicz:2015mha} & 65 / 70& 70 / 70& 68 / 70  \\ 
  HERA1+2 NCep 920 \cite{Abramowicz:2015mha} & 414 / 377& 433 / 377& 471 / 377  \\ 
  HERA1+2 NCep 460 \cite{Abramowicz:2015mha} & 221 / 204& 217 / 204& 225 / 204  \\ 
  HERA1+2 NCep 575 \cite{Abramowicz:2015mha} & 216 / 254& 224 / 254& 222 / 254  \\ 
  Correlated $\chi^2$  total\ (charm) & 86\ (10.5)& 91\ (12.5)& 105\ (11.3)  \\ 
  Log penalty $\chi^2$ total\ (charm) & +6.7\ (+0.1) & -0.7\ (-0.4)& -1.2 \ (-0.2)  \\ 
  \rowcolor{white}
      \midrule
  Total $\chi^2$ / dof  & 1386 / 1207& 1430 / 1207& 1479 / 1207  \\ 
        \bottomrule
    \end{tabular}
  \end{center}
\caption{The $\chi^2$ values  at NLO for individual data sets for a selection of
the charm matching scales $\mu_c$. 
The contribution of the charm data contained in the ``Correlated $\chi^2$'' and in the  ``Log penalty $\chi^2$''  terms
is indicated separately in the parentheses.}
\label{tab:charmNLO}
\end{table*}
}
\def\tableCharmNNLO{
\begin{table*}[t]
  \begin{center}
  \rowcolors{2}{lightgray}{}
    \begin{tabular}{lp{1.65cm}p{1.65cm}p{1.65cm}p{1.65cm}p{1.65cm}}
      \toprule
    {\bf Charm NNLO}    & \hspace{-0.35in} $\mu_c = \quad  1\, m_c$   & $2\, m_c$   & $3\, m_c$       \\ 
      \midrule

  Charm cross section H1-ZEUS combined & 45 / 47& 50 / 47& 50 / 47  \\ 
  H1 F2 Beauty Vertex & 3.5 / 12& 3.5 / 12& 3.3 / 12  \\ 
  Beauty cross section ZEUS Vertex & 13 / 17& 13 / 17& 13 / 17  \\ 
  HERA1+2 CCep & 43 / 39& 43 / 39& 43 / 39  \\ 
  HERA1+2 CCem & 55 / 42& 55 / 42& 54 / 42  \\ 
  HERA1+2 NCem & 217 / 159& 217 / 159& 217 / 159  \\ 
  HERA1+2 NCep 820 & 66 / 70& 64 / 70& 66 / 70  \\ 
  HERA1+2 NCep 920 & 444 / 377& 433 / 377& 442 / 377  \\ 
  HERA1+2 NCep 460 & 218 / 204& 219 / 204& 216 / 204  \\ 
  HERA1+2 NCep 575 & 220 / 254& 218 / 254& 219 / 254  \\ 
  Correlated $\chi^2$ total\ (charm)  & 111\ (10.8)& 109\ (11.3)& 110\ (14.5)  \\ 
  Log penalty $\chi^2$ total\ (charm) & +18\ (-1.1)  & +18\ (-1.8) & +15\ (-1.8)  \\ 
  \rowcolor{white}
      \midrule
  Total $\chi^2$ / dof  & 1453 / 1207& 1439 / 1207& 1447 / 1207  \\ 
      \bottomrule

    \end{tabular}
  \end{center}
\caption{The $\chi^2$ values  at NNLO for individual data sets for a selection of
the charm matching scales $\mu_c$. 
The contribution of the charm data contained in the ``Correlated $\chi^2$'' and in the  ``Log penalty $\chi^2$'' terms 
is indicated separately in the parentheses.
}
\label{tab:charmNNLO}
\end{table*}
} %
\def\tableBottomNLO{
\begin{table*}[t]
  \begin{center}
  \rowcolors{2}{lightgray}{}
    \begin{tabular}{lp{1.65cm}p{1.65cm}p{1.65cm}p{1.65cm}p{1.65cm}}
      \toprule
    {\bf Bottom NLO}    & \hspace{-0.35in} $\mu_b = \quad  1\, m_b$   & $3\, m_b$   & $5\, m_b$   & $10\, m_b$   & $14\, m_b$     \\ 
      \midrule

  Charm cross section H1-ZEUS combined & 46 / 47& 46 / 47& 46 / 47& 46 / 47& 46 / 47  \\ 
  H1 F2 Beauty Vertex & 3.1 / 12& 3.2 / 12& 3.1 / 12& 3.2 / 12& 3.2 / 12  \\ 
  Beauty cross section ZEUS Vertex & 12 / 17& 12 / 17& 12 / 17& 12 / 17& 14 / 17  \\ 
  HERA1+2 CCep & 44 / 39& 44 / 39& 44 / 39& 44 / 39& 44 / 39  \\ 
  HERA1+2 CCem & 52 / 42& 52 / 42& 52 / 42& 53 / 42& 53 / 42  \\ 
  HERA1+2 NCem & 220 / 159& 219 / 159& 220 / 159& 219 / 159& 219 / 159  \\ 
  HERA1+2 NCep 820 & 65 / 70& 65 / 70& 65 / 70& 65 / 70& 65 / 70  \\ 
  HERA1+2 NCep 920 & 414 / 377& 410 / 377& 410 / 377& 412 / 377& 412 / 377  \\ 
  HERA1+2 NCep 460 & 221 / 204& 221 / 204& 221 / 204& 219 / 204& 220 / 204  \\ 
  HERA1+2 NCep 575 & 216 / 254& 216 / 254& 216 / 254& 216 / 254& 216 / 254  \\ 
  Correlated $\chi^2$  total\ (bottom)& 86\ (0.8) & 86\ (0.8) & 86\ (0.8) & 87\ (0.8) & 89\ (0.8)   \\ 
  Log penalty $\chi^2$ total\ (bottom) & +6.7\ (-0.1)& +4.2\ (-0.1)& +4.5\ (-0.1)& +6.6\ (-0.1)& +7.3\ (-0.1)  \\ 
  \rowcolor{white}
      \midrule
  Total $\chi^2$ / dof  & 1386 / 1207& 1379 / 1207& 1380 / 1207& 1383 / 1207& 1388 / 1207  \\ 
      \bottomrule
    \end{tabular}
  \end{center}
\caption{The $\chi^2$ values  at NLO for individual data sets for a selection of
the bottom matching scales $\mu_b$. 
The contribution of the bottom data contained in the ``Correlated $\chi^2$'' and in the  ``Log penalty $\chi^2$''  terms
is indicated separately in the parentheses.
}
\label{tab:bottomNLO}
\end{table*}
} %
\def\tableBottomNNLO{
\begin{table*}[t]
  \begin{center}
  \rowcolors{2}{lightgray}{}
    \begin{tabular}{lp{1.65cm}p{1.65cm}p{1.65cm}p{1.65cm}p{1.65cm}}
      \toprule
    {\bf Bottom NNLO}   & \hspace{-0.35in} $\mu_b = \quad  1\, m_b$   & $3\, m_b$   & $5\, m_b$   & $10\, m_b$   & $14\, m_b$     \\ 
      \midrule

  Charm cross section H1-ZEUS combined & 45 / 47& 45 / 47& 45 / 47& 45 / 47& 45 / 47  \\ 
  H1 F2 Beauty Vertex & 3.5 / 12& 3.7 / 12& 3.7 / 12& 3.6 / 12& 3.6 / 12  \\ 
  Beauty cross section ZEUS Vertex & 13 / 17& 13 / 17& 13 / 17& 13 / 17& 14 / 17  \\ 
  HERA1+2 CCep & 43 / 39& 43 / 39& 43 / 39& 42 / 39& 42 / 39  \\ 
  HERA1+2 CCem & 55 / 42& 55 / 42& 55 / 42& 55 / 42& 56 / 42  \\ 
  HERA1+2 NCem & 217 / 159& 216 / 159& 220 / 159& 218 / 159& 218 / 159  \\ 
  HERA1+2 NCep 820 & 66 / 70& 66 / 70& 66 / 70& 66 / 70& 66 / 70  \\ 
  HERA1+2 NCep 920 & 444 / 377& 445 / 377& 445 / 377& 451 / 377& 453 / 377  \\ 
  HERA1+2 NCep 460 & 218 / 204& 219 / 204& 219 / 204& 217 / 204& 218 / 204  \\ 
  HERA1+2 NCep 575 & 220 / 254& 219 / 254& 219 / 254& 219 / 254& 219 / 254  \\ 
  Correlated $\chi^2$  total\ (bottom) & 111\ (0.9) & 112\ (0.9) & 112\ (0.9) & 114\ (0.9) & 116\ (0.9)   \\ 
  Log penalty $\chi^2$  & +18& +17& +15& +18& +18  \\ 
  \rowcolor{white}
      \midrule
  Total $\chi^2$ / dof  & 1453 / 1207& 1453 / 1207& 1457 / 1207& 1463 / 1207& 1470 / 1207  \\ 
      \bottomrule
    \end{tabular}
  \end{center}
\caption{The $\chi^2$ values at NNLO for individual data sets for a selection of
the bottom matching scales $\mu_b$. 
The contribution of the bottom data contained in the ``Correlated $\chi^2$'' and in the  ``Log penalty $\chi^2$''  terms
is indicated separately in the parentheses.
}
\label{tab:bottomNNLO}
\end{table*}
} %

%% file: text.tex
\newpage

\section{Introduction}
\figvfnsi

The global analyses of PDFs has progressed significantly in recent
years. On the experimental front, there is data ranging from the
fixed-target regime at low energy, on to HERA and the LHC at very high
energies. On the theoretical front, the analysis can be performed not
only at NLO, but now at NNLO.
To capitalize on these advances, it is essential to include a proper
treatment of the heavy quarks to enable 
high precision  phenomenological analysis of measurements.

The Variable Flavor Number Scheme (VFNS) allows us to deal with the
heavy quark mass scale across the full kinematic range by varying the
number of active flavors ($N_F$) in the DGLAP QCD
evolution~\cite{Aivazis:1990pe,Aivazis:1993kh,Aivazis:1993pi,Thorne:2000zd,Martin:2010db,Forte:2010ta,Ball:2011mu,Alekhin:2009ni,Alekhin:2012ig,Alekhin:2013nda,Stavreva:2012bs}.
At low energy scales, the DGLAP evolution only involves $N_F$ light
flavors, and there is no PDF for the heavy quark.
At high energy, the heavy quark PDF is included in the DGLAP evolution
so that there are now $N_F+1$ active flavors.
To combine the above $N_F$ and $N_F+1$ sub-schemes into a single VFNS,
we must define an energy scale $\mu_m$ where we match these together;
this will be the scale where we introduce the heavy quark PDF.

Historically, the matching scale $\mu_m$ was taken to be the heavy
quark mass $m_H$.
At the matching scale, the PDFs and $\alpha_S(\mu)$ for $N_F+1$ are
defined in terms of the $N_F$ quantities by the following boundary
conditions:
\begin{eqnarray}
f_i^{(N_F+1)}(x,\mu_m)&=& \sum_{j} \ {\cal M}_{i}^{j} \ \otimes \ f_j^{(N_F)}(x,\mu_m) 
\label{eq:matchi}
\\
\alpha_S^{(N_F+1)}(\mu_m)&=& \alpha_S^{(N_F)}(\mu_m)
\ \times
\nonumber \\
&&\hspace{-0.3in}
\times\ 
\left( 1+
\sum_{n=1}^{\infty} \sum_{k=0}^{n} c_{n \, k}
\left[
 \alpha_S^{(N_F)}(\mu_m)
\right]^n \ 
\ln^k \frac{\mu_m^2}{m_H^2}
\right) \ .
\label{eq:matchii}
\end{eqnarray}
The matching matrix ${\cal M}_{i}^{j}$ and coefficients $c_{n \, k}$
can be perturbatively computed.\footnote{The perturbative coefficients
  of ${\cal M}_{i}^{j}$ at NLO are available in
  Refs.~\cite{Qian:1985xp,Collins:1986mp}, and at NNLO in
  Ref.~\cite{Buza:1995ie}.
  $m_H$ is the mass of the $N_F+1$ flavor quark.
  For $\alpha_S(\mu)$, the $c_{n \, k}$ coefficients are available in
  the Particle Data Group review of Quantum
  Chromodynamics~\cite{Olive:2016xmw}.  }

The new {\tt xFitter}~2.0.0 program\footnote{Information on the {\tt
    xFitter} program can be found at {\tt www.xFitter.org}, and in
  Refs.~\cite{Alekhin:2014irh,Zenaiev:2016jnq}.  }  links to the {\tt
  APFEL} code~\cite{Bertone:2013vaa} which has implemented generalized
matching conditions that enable the switch from $N_F$ to $N_F+1$
at an arbitrary matching scale $\mu_m$.
This allows us to introduce the heavy quark PDF at any scale---not
just at $\mu_m=m_H$; this flexibility provides a number of advantages.
For example, as the matching scale moves to higher scales, the theory
at the lower scales effectively becomes a Fixed Flavor Number Scheme
(FFNS); yet we still retain a VFNS at the higher scales.

The choice of the matching scale $\mu_m$, like the choice of VFNS or
FFNS, amounts to a theoretical scheme choice.  As such, the variation
of $\mu_m$ represents a source of theoretical uncertainty.
The variable matching scale implemented in {\tt xFitter} provides a
new incisive tool to study the impact of these choices across a broad
kinematic region.
Additionally, as we move from NLO to NNLO calculations, new features
are encountered, and these compel us to reexamine some of the
foundational elements used to construct this theoretical framework.

Reconsidering the historical choice $\mu_m = m_H$ is of particular
relevance for heavy-quark initiated processes at the LHC. In this
context, the benefits of the FFNS close to the threshold region and of
the VFNS at higher scales are often simultaneously needed to describe
the data. Therefore, a careful choice of the matching scales could
help formulate a matching prescription between FFNS and VFNS able to
achieve this goal in a very simple fashion~\cite{Bertone:2017inprep}.

This study will examine the combined HERA data set and evaluate the
impact of the matching scale on the features of the fit of PDFs.
In Sect.~\ref{sec:vfns}, we review the key elements of the VFNS used
in this study.
Sect.~\ref{sec:match}, shows the impact of the matching scale $\mu_m$
on the PDFs.
In Sect.~\ref{sec:fit}, we perform a fit of the combined HERA data
sets at both NLO and NNLO, and investigate the effect of the matching
scale $\mu_m$.
Sect.~\ref{sec:example} presents an example of how the $\mu_m$
flexibility can be used as a tool to evaluate a recent suggestion for
a $N_F$ dependent PDF.
Sect.~\ref{sec:conclusion} summarizes the general characteristics and
conclusions of this study.

\section{Variable Flavor Number Scheme (VFNS)}
\label{sec:vfns}

Here we will outline the key concepts of the heavy quark VFNS which
are relevant for this investigation.

\subsection{The matching scale $\mu_{m}$}

A generalized formulation of the VFNS factorization is based on the
Collins-Wilczek-Zee (CWZ) renormalization scheme which involves a
sequence of sub-schemes parameterized by the number of active quark
flavors ($N_{F}$)~\cite{Collins:1978wz,Collins:1998rz}.
For each sub-scheme, the $N_F$ (active) flavors are renormalized using
the $\overline{\mbox{MS}}$ scheme while the heavy (inactive) flavors are
renormalized using zero-momentum subtraction.  This ensures that to
all orders in perturbation theory (i)~the results are gauge invariant,
(ii)~the results for the active $N_F$ flavors match the standard
$\overline{\mbox{MS}}$ results, and (iii)~the heavy (inactive) flavors
manifestly decouple.\footnote{%
  For the CWZ scheme with $N_F$ (active) flavors and an arbitrary
  number of heavy (inactive) flavors, the evolution of the PDFs and
  $\alpha_S^{(N_F)}(\mu)$ will involve only the active $N_F$ flavors;
  the inactive heavy flavors can be ignored.  }
Specifically, both the DGLAP evolution kernels for the $N_F$ active
PDFs and the renormalization group equation for
$\alpha_S^{(N_F)}(\mu)$ are pure $\overline{\mbox{MS}}$.

To connect the separate $N_F$ sub-schemes into a single scheme that
spans the full kinematic range, we must choose a matching scale
$\mu_{m}$ which will relate the sub-schemes.
This is where we define the PDFs and $\alpha_S$ of the $N_F+1$ scheme
in terms of the $N_F$ scheme, {\it cf.} Eqs.~(\ref{eq:matchi})
and~(\ref{eq:matchii}).
A schematic representation of this is displayed in
Fig.~\ref{fig:vfns1}.

For example, at scales $\mu_c< \mu <\mu_b$ the scheme has $N_F=4$
active flavors $\{u,d,s,c\}$ with 4-flavor PDFs and
$\alpha_S^{(4)}(\mu)$; the bottom quark is {\bf not} treated as a
parton and $f_b^{(4)}(x,\mu)=0$.

At the scale $\mu =\mu_b$, we can compute the 5-flavor PDFs and
$\alpha_S^{(5)}(\mu)$ in terms of the 4-flavor quantities; the
boundary conditions are non-trivial and the PDFs and $\alpha_S(\mu)$
are not necessarily continuous.  This scheme has $N_F=5$ active
flavors $\{u,d,s,c,b\}$, and the bottom quark is included in the DGLAP
evolution.

\subsection{Historical choice of $\mu_{m}=m_{c,b,t}$}

Historically, the matching scale $\mu_{m}$ was commonly taken to be
exactly equal to the mass of the heavy quark $\mu_{m}=m_{c,b,t}$; this
was a convenient choice for a number of reasons.

For example, the generic NLO matching condition for the PDFs at the
$N_F=4$ to $N_F=5$ transition is~\cite{Kusina:2013slm}:
\begin{equation}
  f_{i}^{(5)}(x,\mu_b) =\left\{
  \delta_{ij} + 
  \frac{\alpha_{S}^{(4)}(\mu_b)}{2\pi}
  \left[c_{0}^{ij}+c_{1}^{ij} \ln \left(\frac{\mu_b^2}{m_b^2}  \right)\right]\right\}
  \otimes f_{j}^{(4)}(x,\mu_b) 
\label{eq:match}
\end{equation}
where $c_{0}^{ij}$ and $c_{1}^{ij}$ are perturbatively calculable
coefficient functions.  Note that the right-hand side uses 4-flavor
PDFs and $\alpha_S$, while the left-hand side uses 5-flavors.

The choice $\mu_{b}=m_{b}$ will cause the logarithms to vanish, and
this greatly simplifies the matching relations.
Additionally, at NLO in the $\overline{\mbox{MS}}$ scheme the constant term
$c_{0}^{ij}$ in the matching equation coincidentally
vanishes~\cite{Buza:1995ie}.
The net result is that for $\mu_{b}=m_{b}$, the PDFs will be
continuous (but not differentiable) at NLO.
This is historically why $\mu_m$ was set to $m_{c,b,t}$.

However, at NNLO and beyond the situation is more complex; in
particular, the higher-order terms corresponding to $c_{0}^{ij}$ will
be non-zero, and the matching of both the PDFs and $\alpha_S(\mu)$
will be discontinuous. Consequently, the freedom to arbitrarily choose
the matching scale $\mu_{m}$ (and decide where to place the
discontinuities) will have a number of advantages, as the next
subsection will demonstrate.

\subsection{Smooth matching across flavor thresholds \label{subsec:matching}}
\figacot

To gauge the impact of the contributions of the heavy quark PDFs in a
process independent manner, we can compare the DGLAP evolved heavy
quark PDF $f_b(x,\mu)$ with a perturbatively computed quantity:
$\widetilde{f}_b(x,\mu)$.
At NLO, $\widetilde{f}_b(x,\mu)$ takes a gluon PDF and convolutes it
with a perturbative (DGLAP) splitting
$g\to b \bar{b}$~\cite{Maltoni:2012pa,Lim:2016wjo}; this can be
thought of as a ``perturbatively'' computed bottom PDF.
The result at NLO is:
\begin{equation}
  \widetilde{f}_b(x,\mu) = 
  \frac{\alpha_S}{2 \pi}\ P_{g\to b \bar{b}}\otimes f_g \ \  \ln \left[ \frac{\mu^2}{m_b^2} \right] \quad .
  \label{eq:ftilde}
\end{equation}
The difference between $f_b(x,\mu)$ and $\widetilde{f}_b(x,\mu)$ is
due to the higher order terms which are resummed by the heavy quark
DGLAP evolution.\footnote{%
  In Eq.~(\ref{eq:ftilde}), $\widetilde{f}_b(x,\mu)$ includes the
  single splitting $(g\to b \bar{b})$; in contrast, the DGLAP
  evolution of $f_b(x,\mu)$ sums an infinite tower of splittings.
  Note, we have used the \hbox{NNPDF30\_lo\_as\_118\_nf\_6} PDFs to
  precisely match the order of the splitting kernels in the NLO
  calculation.}

To better understand these quantities, we compute DIS bottom
production at NLO in a 5-flavor VFNS, and find the cross section to
be~\cite{Aivazis:1993pi}:
\begin{equation}
  \sigma_{\scriptscriptstyle{VFNS}}=
  \sigma_{b\to b}\otimes 
  \left[
    f_b(x,\mu) -   \widetilde{f}_b(x,\mu) 
  \right]
  +
  \underbrace{
    \sigma_{g\to b}\otimes f_g(x,\mu)
  }_{\displaystyle{\sim \sigma_{\scriptscriptstyle{FFNS}}}}
  \quad . 
\end{equation}
Here, $\sigma_{b\to b} \otimes f_b$ is the LO term, and
$\sigma_{b\to b} \otimes \widetilde{f}_b$ is the subtraction (SUB)
term.
The unsubtracted NLO term $\sigma_{g\to b}\otimes f_g$ corresponds
(approximately) to a FFNS calculation.
Here, $\sigma_{b \to b}$ is proportional to a delta function
which makes the convolution trivial.

Thus, the combination $(f_b - \widetilde{f}_b)$ represents
(approximately) the difference between a VFNS and FFNS
result.\footnote{%
The above correspondences are only approximate as the
  VFNS and FFNS also differ in $\alpha_S$ and the PDFs.
}
These quantities are displayed in Fig.~\ref{fig:acot}.
In the region $\mu\sim m_b$, $f_b(x,\mu)$ and $\widetilde{f}_b(x,\mu)$
match precisely; it is this cancellation which (at NLO) ensures
physical quantities will have a smooth transition across the flavor
threshold.

At larger $\mu$ scales, $f_b(x,\mu)$ and $\widetilde{f}_b(x,\mu)$
begin to diverge; this indicates that the resummed heavy quark
logarithms are becoming sizable.  The details clearly depend on the
specific $x$ values. For large $x$ ($x\sim 0.1$) we find
$f_b(x,\mu) >\widetilde{f}_b(x,\mu)$, while for small $x$
($x\sim 0.001$) the result is $f_b(x,\mu) <\widetilde{f}_b(x,\mu)$;
finally, for intermediate $x$ ($x\sim 0.01$) the two terms nearly
balance even for sizable $\mu$ scales.

While the QCD theory ensures proper matching, this is not so easy to
implement in a general numeric calculation for all observables,
especially for complex observables involving multiple numeric
integrations. In particular, the cancellation of Fig.~\ref{fig:acot}
requires that the quark masses $m_{c,b,t}$, the strong coupling
$\alpha_{S}$, and the order of the PDF evolution are exactly matched
in (i)~the DGLAP evolution that generates the PDFs, (ii)~the partonic
cross sections that are convoluted with the PDFs, and (iii)~the
fragmentation function (if used).

In practice, there are almost always slight differences.  A typical
analysis might use a variety of PDFs from different PDF groups,
together with a selection of fragmentation functions; each of these
will be generated with a specific set of quark masses and $\alpha_S$
values which are most likely different.
Thus, it is essentially inevitable that the cancellations exhibited in
Fig.~\ref{fig:acot} will be spoiled leading to spurious contributions
which can be substantive.

Instead of setting the matching scale at the heavy quark mass
$\mu_m=m_{c,b,t}$, {\tt xFitter} provides the flexibility to delay the
matching scale $\mu_{m}$ to a few multiples of the heavy quark mass;
this will avoid the need for the delicate cancellation in the
$\mu_m\sim m_{c,b,t}$ region, and the results will be numerically more
stable.

As an extreme example, one could imagine delaying the matching scale
to infinity ($\mu_{m}\to\infty$) which would amount to a FFNS; here,
the disadvantage is that the FFNS does not include the resummation of
the higher-order heavy quark logs which have been demonstrated to
improve the fit to the data~\cite{Ball:2013gsa}. Using the new
flexibility of the {\tt xFitter} program, it is possible to
investigate the trade-offs between a large and small value for the
matching scale $\mu_{m}$.

A separate example is present in the transverse momentum ($p_T$)
distributions for heavy quark production ($pp\to b \bar{b}$) using the
(general mass) GM-VFNS~\cite{Cacciari:1998it,Kniehl:2015fla}. If we compute this in an
$N_F=5$ flavor scheme, the contribution from the
$b\bar{b}\to b\bar{b}$ sub-process with an exchanged $t$-channel gluon
will be singular at $p_T=0$.  For a scale choice of  the transverse
mass $\mu=\sqrt{p_T^2+m_b^2}$ (a common choice), the singularity can
be cured by either a different scale choice, or by delaying the switch
to the 5-flavor scheme to a higher scale, {\it e.g.},
$\mu_b\sim 2 m_b$.

\subsection{Discontinuities}

At NNLO both the PDFs and the $\alpha_{S}(\mu)$ will necessarily have
discontinuities when matching between the $N_{F}$ to $N_{F}+1$ flavor
schemes as specified by Eqs.~(\ref{eq:matchi}) and~(\ref{eq:matchii}).
If we are analyzing a high precision experiment and arbitrarily impose
a matching at the quark masses $\mu_{m}=m_{c,b,t}$, this may well
introduce discontinuities within the kinematic range of some precision
data. While it is true that these discontinuities simply reflect the
theoretical uncertainties, it is disconcerting to insert them in the
middle of a precision data set.

The ability to vary the matching scale $\mu_{m}$ provides us with the
option to shift the location of these discontinuities for a particular
analysis.
For example, to analyze the high-precision charm production HERA data,
we necessarily are working in the region of the bottom mass scale
($\sim4.5$~GeV).
Both the PDFs and $\alpha_{S}(\mu)$ will be discontinuous at the
matching scale which transitions between the $N_{F}=4$ and $N_{F}=5$
schemes.
If the matching scale is chosen in the region $\mu_{m}\sim m_{b}$,
these discontinuity will appear in the region of the data.
Instead, we can shift the matching $\mu_{m}$ to a higher scale (for
example, set $\mu_{m}$ to $2m_{b}$ or $3m_{b}$) and thus analyze the
charm production data in a consistent $N_{F}=4$ flavor framework. Yet,
we still retain the transition to $N_{F}=5$ flavors so that processes
such as LHC data at high scales are computed including the bottom PDF.

\section{The matching scale $\mu_m$ }
\label{sec:match}

Having sketched the characteristics of a flexible matching scale
$\mu_{m}$, we will examine the specific boundary condition and the
impact on the global fit of the PDFs.

\subsection{Impact of matching on the PDFs}
\figBmatch %
\figFbottom %

Fig.~\ref{fig:bMatch} displays the effect of different values of the
bottom matching scale $\mu_{b}$ on the bottom-quark PDF for both the
NLO and NNLO cases.\footnote{
  A first study of the impact of moving the bottom matching scale with
  respect to the bottom mass was already done in
  Ref.~\cite{Bonvini:2015pxa} in the context of $\overline{b}bH$
  production at the LHC using a matched scheme. The approach developed
  in this study was more recently applied to the 13 TeV LHC in
  Ref.~\cite{Bonvini:2016fgf}.
}
At NLO, the matching conditions are schematically:\footnote{
  At NNLO, the bottom-quark matching condition also receives
  contributions from the light quarks as well as gluons; this has been
  included in the calculation. }
\begin{equation}
  f_{b}^{(5)}(x,\mu_b)=
  \frac{\alpha_{S}^{(4)}(\mu_b)}{2\pi} \ 
  \left[c_{0}^{bg}+c_{1}^{bg}\ L\right]
  \otimes f_{g}^{(4)}(x,\mu_b)
\label{eq:bMatch}
\end{equation}
where $L=\ln(\mu_b^{2}/m_{b}^{2})$. The superscripts $\{4,5\}$
identify the number of active flavors $N_{F}$. The gluon and the light
quarks also have matching conditions analogous to
Eq.~(\ref{eq:bMatch}).

As already mentioned, if we choose to match at $\mu_b=m_{b}$ then $L=0$ and
$f_{b}^{(5)}(x,\mu_m)$ will start from zero at $\mu_b=m_{b}$. This
coincidental zero ($c_{0}^{ij}=0$) is the historic reason why most NLO
analyses perform the matching at $\mu_b=m_{b}$; if both the
$c_{0}^{ij}$ and $c_{1}^{ij}\, L$ terms can be ignored, then the PDFs
are continuous (but not differentiable) across the matching
scale.\footnote{
  While the VFNS framework is compatible with an intrinsic charm or
  bottom PDF, we do not introduce these into the current study.  For
  additional details, see
  Refs.~\cite{Ball:2016neh,Ball:2015tna,Ball:2015dpa,Lyonnet:2015dca}.}

At NNLO this is no longer the case; the NNLO constant term at
${\cal O}(\alpha_{S}^{2})$ does not vanish and the PDFs will have a
discontinuity regardless of the choice of matching scale. Although the
difference is subtle, the (red) curve for $\mu_{b}=m_{b}$ does start
exactly from zero for the NLO calculation (Fig.~\ref{fig:bMatch}-a),
while for the NNLO calculation (Fig.~\ref{fig:bMatch}-b) it starts
from a small non-zero value.

As we vary the matching $\mu_{b}$ in the vicinity of $m_{b}$, the sign
of $f_{b}^{(5)}(x,\mu_b)$ is controlled by the log term
$(c_{1}^{ij}\,L)$. For $\mu_b<m_{b}$ this combination will drive
$f_{b}^{(5)}(x,\mu_b)$ negative, and this will be compensated (in the
sum rule for example) by a positive shift in the 5-flavor gluon. Thus,
QCD ensures that both momentum and number sum rules are satisfied to
the appropriate order.

Comparing different $f_{b}^{(5)}(x,\mu)$ curves computed with the NLO
matching conditions (Fig.~\ref{fig:bMatch}-a) at large $\mu$ scales,
there are obvious differences in the curves. This reflects the
difference between the single log contribution $(c_{1}^{ij}\, L)$
computed by the matching condition of Eq.~(\ref{eq:bMatch}) and the
resummed contributions computed by the DGLAP evolution
equation. Specifically, the NLO matching includes the $\alpha_{S}L$
contribution, but is missing $\alpha_{S}^{2}L^{2}$ and higher terms;
this is what gives rise to the differences of
Fig.~\ref{fig:bMatch}-a. Obviously, the $\alpha_{S}^{2}L^{2}$
contributions can be important.

Comparing the different $f_{b}^{(5)}(x,\mu)$ curves computed with the
NNLO matching conditions (Fig.~\ref{fig:bMatch}-b) at large $\mu$
scales, the differences in the curves are greatly reduced compared to
the NLO case. The NNLO result includes both the $\alpha_{S}L$ and
$\alpha_{S}^{2}L^{2}$ contributions, but is missing
$\alpha_{S}^{3}L^{3}$ and higher orders. Clearly the inclusion of the
$\alpha_{S}^{2}L^{2}$ contributions dramatically reduces the effect of
the different choices of the $\mu_{m}$ matching scale.

Finally, we wish to emphasize that ultimately the choice of $\mu_{m}$
amounts to a choice of scheme. In the limit that perturbation theory
is computed to all orders, the infinite tower of logarithms resummed
by the DGLAP evolution equations (in the $N_{F}+1$-flavor scheme) will
be explicitly summed in the matching conditions (in the $N_{F}$-flavor
scheme). In a practical sense, while the differences at NLO are
substantive, at NNLO the residual differences at large $\mu$ scale are
much smaller.
This reduced sensitivity on the choice of $\mu_{m}$ provides increased
flexibility and precision in our fits, as will be illustrated in the
following sections.

\subsection{Impact of matching on $F_2^b(x,Q)$}

Having examined the PDFs in the previous section we now turn to a
physical observable, $F_2^b(x,Q)$.

Fig.~\ref{fig:fbottom}-a) shows the NLO result for $F_2^b(x,Q)$ which
will receive contributions from the LO process ($\gamma b\to b$) as
well as the NLO ($\gamma g \to b \bar{b}$) process.
For $\mu<\mu_b$, $f_b^{(5)}(x,\mu)=0$ and only the gluon initiated
process contributes.
For $\mu\gtrsim \mu_b$, the bottom PDF turns on ({\it cf.}
Fig.~\ref{fig:bMatch}), and the heavy quark initiated process now
contributes.  Because the PDFs, $\alpha_S(\mu)$, and $m_b$ are all
carefully matched in this calculation, the cancellation outlined in
Sect.~\ref{subsec:matching} ensures that the prediction for the
physical observable is relatively smooth in this region.

Fig.~\ref{fig:fbottom}-b) shows the NNLO result for $F_2^b(x,Q)$.  As
with the PDF matching of Fig.~\ref{fig:bMatch}-b), the additional NNLO
contributions significantly reduce the impact of the different
matching scales so that the prediction for $F_2^b(x,Q)$ is now very
insensitive to $\mu_b$.

The above smooth transition of $F_2^b(x,Q)$ from the $N_F=4$ to the
$N_F=5$ scheme holds even though the PDFs and $\alpha_S(\mu)$ have
discontinuities.
Because we have used consistent choices for
$\{ m_b, f_i^{(N_F)}, \alpha_S \}$, the cancellation of
Sect.~\ref{subsec:matching} applies, and the effect of any
discontinuities in the physical observable will be of higher order.
Conversely, a mismatch in $\{ m_b, f_i^{(N_F)}, \alpha_S \}$ would
spoil this cancellation and result in unphysical large contributions
when $f_b^{(5)}(x,\mu)$ is introduced. This is precisely the case
where shifting the matching scale $\mu_b$ to a higher scale such as
$2m_b$ or $3m_b$ would help avoid these problems.

It is interesting to note that as we compute even higher orders, the
discontinuities in the PDFs and $\alpha_S(\mu)$ will persist at lower
order; but, any discontinuities in the physical observables will
systematically decrease order by order.

\section{The PDF fits}
\label{sec:fit}

\subsection{{\tt xFitter}, {\tt APFEL}, and data sets }

To study the effects of varying the matching scales for the charm and
bottom quark we will perform a series of fits to various data
sets. Since we are varying the matching scales in the vicinity of
$m_{c}$ and $m_{b}$, we want data that constrain the PDFs in this
region. For this purpose, we include the very precise combined HERA
data sets as these provide strong constraints in the region
$\mu\sim m_{c,b}$, and also extend up to higher
scales~\cite{Abramowicz:2015mha,Aaron:2009af,Abramowicz:1900rp,Abramowicz:2014zub}.
In particular, the HERA measurement of the charm and bottom cross
sections are included as they are sensitive to the choice of $\mu_c$
and $\mu_{b}$.

These fits are performed with the {\tt xFitter} program
using the {\tt APFEL} evolution
code~\cite{Bertone:2013vaa,Carrazza:2014gfa,Bertone:2016lga}. The DIS
calculations use the FONLL-B scheme for the NLO calculations, and the
FONLL-C scheme for the NNLO calculations;
these are both  ${\cal O}(\alpha_S^2)$ 
prescriptions, and the details are specified in Ref.~\cite{Forte:2010ta}.
We use $m_c=1.45$~GeV, $m_b=4.5$~GeV, $\alpha_S(M_Z)=0.118$ for both
the NLO and NNLO calculations.
The fit is performed using pole masses, but the formalism can be used
equally well with the $\overline{\mbox{MS}}$ definition of the heavy
quark masses~\cite{Bertone:2016ywq}.
For the PDFs, we use a HERAPDF 14-parameter functional form with 
initial QCD evolution scale 
$Q_0^2=1.0~{\rm GeV}^2$ and strangeness fraction  $f_s=0.4$; 
the other QCD fit settings and constraints are similar  to the analysis of Ref.~\cite{Bertone:2016ywq}.

The minimization of the $\chi^2$ is performed using
MINUIT~\cite{James:1975dr}.  The correlations between data points
caused by systematic uncertainties are taken into account in the
``Correlated $\chi^2$'' contribution.  A ``Log penalty $\chi^2$''
arises from the likelihood transition to $\chi^2$ when the scaling of
the errors is applied~\cite{Aaron:2012qi,Alekhin:2014irh}.

The full sets of data are listed in
Tabs.~\ref{tab:charmNLO}-\ref{tab:bottomNNLO}, and the reference for
each data set is cited in Tab.~\ref{tab:charmNLO}.
The combined inclusive HERA data (HERA1+2) from
Ref.~\cite{Abramowicz:2015mha} includes both neutral current (NC) and
charged current (CC) results for electrons (em) and positrons (ep) at
a variety of energies.
The charm cross sections from Ref.~\cite{Abramowicz:1900rp} include
the combined H1-ZEUS results.
The bottom cross sections from ZEUS are presented in
Ref.~\cite{Abramowicz:2014zub} and those from H1 in
Ref.~\cite{Aaron:2009af}.

\subsection{Impact of matching on the fits: charm}
\figCharm
\figCharmOnly
\tableCharmNLO
\tableCharmNNLO

The charm cross section data are expected to be sensitive to the
treatment of the charm PDF in the threshold region, and this is
reflected in the results of 
Figures~\ref{fig:charm},~\ref{fig:charmOnly} and  
Tables~\ref{tab:charmNLO},~\ref{tab:charmNNLO}.

Fig.~\ref{fig:charm} displays the results for varying the charm quark
matching scale $\mu_{c}$ both for the NLO and NNLO
calculations.\footnote{
  For these scans we hold the bottom matching fixed at $\mu_{b}=m_{b}$
  and keep $\mu_{c}<m_{b}$ so the ordering of the mass thresholds is
  not inverted.}
Comparing the NLO and NNLO cases, the NLO result ranges over $\sim100$
units in $\chi^{2}$, while the NNLO varies over $\sim25$ units of
$\chi^{2}$.
This difference in the $\chi^2$ variation reflects the effects of the
higher order terms; it is reassuring to see that the $\mu_c$
dependence decreases at higher orders.

At NLO, the matching conditions pick up the contribution of only the
single log term $L$ (Eq.~(\ref{eq:bMatch})), while at NNLO we pick up
both the $L$ and $L^2$ terms.  In contrast, the DGLAP evolved charm
PDF resums the above, as well as an infinite tower of logs:
$\sum_{n=1}^{\infty} \sum_{k=0}^{n} \alpha_S^n \, L^k$.

Examining the NLO analysis of Fig.~\ref{fig:charm}-a, we find that at
low scales, the $\chi^2$ increases with increasing $\mu_c$
scale. While our plot extends slightly below the charm mass, it is not
obvious if there is actually a minimum in $\mu_c$.  It is problematic
to compute with $\mu_c$ values much lower than $m_c$ as $\alpha_S$
becomes large and the charm PDF negative.
Thus, the optimal computational range for $\mu_c$ appears to be in the
region of $m_c$.

Focusing on the charm data alone as shown in
Fig.~\ref{fig:charmOnly}-a, the situation is not so clear; the
$\chi^2$ increases with increasing $\mu_c$, but again there does not
appear to be a minimum at low $\mu_c$ values.  Moving to large $\mu_c$,
the $\chi^2$ values initially increase, but then decrease as $\mu_c$
approaches $m_b$.
As we want to maintain the ordering $\mu_c<\mu_b$, we cannot go to
larger scales unless we increase $\mu_b$. While this is allowed, it is
more complex to explore the two-dimensional $\{ \mu_c, \mu_b \}$
parameter space; hence, we limit the present study to variation of a
single scale.

The $\chi^2$ results for each individual data set is summarized in
Tab.~\ref{tab:charmNLO}. The data sets with the largest effects are
i)~the H1-ZEUS combined charm cross section data, and ii)~the very
precise ``HERA1+2NCep 920'' set.
The sensitivity of the ``HERA1+2NCep 920'' set is due to a large number of data points 
with small uncertainties.

Turning to the NNLO analysis of Fig.~\ref{fig:charm}-b and the results 
of Tab.~\ref{tab:charmNNLO}, a number of points are evident. 
Again, the two data sets with the largest impact are the H1-ZEUS
combined charm cross section data, and the ``HERA1+2NCep 920'' set.
In Fig.~\ref{fig:charm} the vertical lines indicate the bin boundaries
for the ``HERA1+2NCep 920'' data set.

Scanning in $\chi^2$, discrete jumps are evident.
As we vary the matching scale, certain data bins move
between the $N_{F}=3$ and $N_{F}=4$ schemes, 
shifting the $\chi^2$ by one or two units which is visible in Fig.~\ref{fig:charm}-b).
These jumps reflect the underlying theoretical uncertainty arising
from the choice of $N_F$.

In Fig.~\ref{fig:charm}-b the total NNLO variation of $\chi^2$ is
reduced compared to the NLO case, and the minimum global $\chi^2$ is now in
the region $\mu_c \sim 2 m_c$.
Focusing on the charm data alone in Fig.~\ref{fig:charmOnly}-b, again
it is not obvious if there is actually a minimum in $\mu_c$.  Given
the limitations of computing with $\mu_c \ll m_c$, the optimal
computational range again appears to be in the general region of
$m_c$.

While it may be tempting to try and optimize the matching scale for
each data set, recall that $\mu_{m}$ represents a choice of scheme,
and thus reflects an inherent theoretical uncertainty; a specific
choice of $\mu_{m}$ will not reduce this uncertainty.

This situation can also be found in complex global fits where the
final result may be a compromise of data sets which are in tension;
this is why a tolerance factor is often introduced.
This complexity is evident when examining the details of 
Tables~\ref{tab:charmNLO} and~\ref{tab:charmNNLO}
which demonstrate the minimum $\chi^2$ for individual data sets
is not simply correlated; 
this will be discussed further in Section~\ref{sec:comparisons}. 
An additional 
challenge of analyzing the charm case is that $\mu_c$ can only vary
over the limited dynamic range between $\sim m_c$ and $\mu_b$. This
will not be an issue for the bottom quark (because $m_t \gg m_b$),
which is considered in the following section.

\subsection{Impact of matching on the fits: bottom}
\figBottom
\figBottomOnly
\tableBottomNLO
\tableBottomNNLO

Fig.~\ref{fig:bottom} presents the results for varying the bottom
quark matching scale $\mu_{b}$ both for the NLO and NNLO
calculations. This figure highlights the  ranges of $\chi^2$;
the NLO result ranges over approximately $\sim 10$ units in $\chi^{2}$, 
and the NNLO varies by about the same amount. 

The reduced  $\chi^{2}$ variation as compared to the charm case 
reflects, in part, the decrease in the strong coupling
$\alpha_{S}(m_{b})<\alpha_{S}(m_{c})$ which also diminishes the 
higher order contributions. 
Fig.~\ref{fig:charm} with Fig.~\ref{fig:bottom} there is a $\chi^{2}$
range of $\sim 100$ vs. $\sim 10$ for NLO, and $\sim 15$ vs. $\sim 10$
for NNLO.

Examining the NLO analysis of Fig.~\ref{fig:bottom}-a, there is a
slight minimum for $\chi^2$ in the region $\mu_b \sim 2 m_b$ with
relatively flat behavior at larger $\mu_b$ scales. Correspondingly,
there is a similar behavior when we focus on only the bottom data of
Fig.~\ref{fig:bottomOnly}-a. The $\chi^2$ results for each individual
data set is summarized in Tab.~\ref{tab:bottomNLO}.

The data sets with the largest effects are i)~the very precise
``HERA1+2NCep 920'' set, and ii)~the separate H1 and ZEUS bottom cross
section data.
The H1 and ZEUS bottom cross sections display some minimal $\chi^2$
variation in the region $\mu_b \sim m_b$, but then is relatively flat
out to very high scales ($\mu_b \sim 14 m_b$).
It is primarily the ``HERA1+2NCep 920'' set which drives the shape of
the $\chi^2$ curve in the $\mu_b\sim m_b$ region.
Compared to the charm results, the interpretation of the bottom cross
section data requires some care as the number of data points is
smaller, and the relative uncertainty larger.

Turning to the NNLO analysis of Fig.~\ref{fig:bottom}-b, the variation
of the $\chi^2$ curve is within $\sim8$ units across the range of the
plot. The resolution of the vertical $\chi^2$ scale accentuates the
discrete jumps as the data bins move between the $N_{F}=4$ and
$N_{F}=5$ schemes. The bin boundaries for the ``HERA1+2NCep 920'' data
set are indicated with vertical lines.

Focusing on the bottom data alone as shown in
Fig.~\ref{fig:bottomOnly}-b, the $\chi^2$ profile is flat within one
unit across the plot range.

For both Fig.~\ref{fig:bottom}-b and Fig.~\ref{fig:bottomOnly}-b, the
$\chi^2$ variation is within a reasonable ``tolerance'' factor for the
global fit; thus, the matching scale $\mu_b$ can vary within this
range with minimal impact on the resulting fit.

The scale $\mu_b$ can extend up to larger scales, and
Tabs.~\ref{tab:bottomNLO} and~\ref{tab:bottomNNLO} display the results
for $10m_b$ and $14m_b$.
The pattern across the various data sets is
consistent, and the overall $\chi^2$ values rise slowly.

\subsection{Comparisons} \label{sec:comparisons}
\figChiScaledi
\figChiScaledii

To facilitate comparisons of the NLO and NNLO results,
Fig.~\ref{fig:chi2scaledi} displays the ratio $\chi^{2}/\chi^{2}_0$
for charm (on the left) and bottom (on the right) where $\chi^{2}_0$
is the value of the $\chi^2$ at $\mu_m=m_H$.
Similarly, Fig.~\ref{fig:chi2scaledii} displays the same ratio for
only the heavy quark data sets.
By plotting $\chi^{2}/\chi^{2}_0$, we can better compare the
fractional variation of $\chi^2$ across the matching scale values.

The motivation for the scaled plot of 
Figs.~\ref{fig:chi2scaledi} and~\ref{fig:chi2scaledii}
is that the overall $\chi^2$ values are different; specifically,
those of the NNLO are greater than the NLO. 
This counter intuitive result has been observed in other analyses~\cite{Ball:2014uwa,Abramowicz:2015mha},
and it has been suggested that this may be improved by resumming  the singular $\ln[1/x]$ terms in the
higher order splitting kernels~\cite{Bonvini:2016wki}.

Here, we first make some observations specific to Figures~\ref{fig:chi2scaledi} and~\ref{fig:chi2scaledii}.

\begin{itemize}
\setlength\itemsep{1em}

\item At NLO for the case of charm, the optimal computational scale
  for $\mu_c$ is in the general range $\mu_c\sim m_c$ for both the
  inclusive data set (Fig.~\ref{fig:chi2scaledi}-a) and the charm data
  set (Fig.~\ref{fig:chi2scaledii}-a).
  For lower scales ($\mu_c \ll m_c$), $\alpha_S(\mu)$ is large and the
  charm PDFs are negative.  For higher scales ($\mu_c \gg m_c$),
  $\chi^2/\chi^2_0$ increases.

\item At NLO for the case of bottom, the optimal scale for $\mu_b$ is
  in the general range $\mu_b\sim 2 m_b$. For the inclusive data set
  (Fig.~\ref{fig:chi2scaledi}-b) the $\chi^2/\chi^2_0$ variation is
  very mild ($\sim 1\%$), while for the bottom data set
  (Fig.~\ref{fig:chi2scaledii}-b) the $\chi^2/\chi^2_0$ variation is
  larger ($\sim 10\%$).

\item At NNLO for the case of charm, the $\chi^2/\chi^2_0$ variation
  is reduced.
  For the inclusive data set (Fig.~\ref{fig:chi2scaledi}-a) the
  $\chi^2/\chi^2_0$ variation is very mild ($\sim 2\%$), while for the
  charm data set (Fig.~\ref{fig:chi2scaledii}-a) the $\chi^2/\chi^2_0$
  variation is larger ($\sim 10\%$).
  There is no obvious optimal choice for the $\mu_c$ scale.

\item At NNLO for the case of bottom, the $\chi^2/\chi^2_0$ variation
  is reduced and a matching scale choice in the region $\mu_b\sim m_b$
  appears to be optimal.
  For the inclusive data set (Fig.~\ref{fig:chi2scaledi}-b) the
  $\chi^2/\chi^2_0$ variation is very mild ($\sim 1\%$), while for the
  bottom data set (Fig.~\ref{fig:chi2scaledii}-b) the
  $\chi^2/\chi^2_0$ variation is slightly larger ($\sim 5\%$).
\end{itemize}

While the detailed characteristics of the above fits will depend on specifics 
of the analysis, there are two general patterns which emerge:
i)~the  $\chi^2$ variation of the NNLO results are generally reduced compared to the NLO results,
and
ii)~the relative $\chi^2$ variation across  the bottom transition is reduced compared to the charm transition. 
For example, although the global $\chi^2$ can be modified by different choices of data sets and
weight factors, these general properties persist  for each individual data set of Tables~\ref{tab:charmNLO}--\ref{tab:bottomNNLO};
in fact,  we see that 
the bulk of the data sets are quite insensitive to the details of the 
heavy quark  matching scale.
Additionally, there are a variety of prescriptions for computing the heavy flavor contributions;
these primarily differ in how the higher order contributions are organized. 
As a cross check, we performed a NLO fit using the  FONNL-A scheme; while
the absolute value of  $\chi^2$ differed, the above general properties persisted.

The net result is that we can now quantify the theoretical uncertainty
associated with the transition between different $N_F$ sub-schemes.
In practical applications, if we choose $\mu_{c}\sim m_c$, the impact
of the $N_F=3$ to $N_F=4$ transition is reduced as this is often below
the minimum kinematic cuts of the analysis ({\it e.g.} $Q_{min}^2$ and
$W_{min}^2$).
Conversely, the $N_F=4$ to $N_F=5$ transition is more likely to fall
in the region of fitted data; hence,  it is useful to quantify the
uncertainty associated with the $\mu_b$ choice.

\section{An example: $N_F$-dependent PDFs} \label{sec:example}

The variable  matching scale $\mu_m$ can be used as an incisive tool to
explore various aspects of the PDFs and global fits.
As an example, Ref.~\cite{Kusina:2013slm} introduced an
$N_F$-dependent PDF  $f_{i}(x,\mu,N_{F})$ where $N_F$ is the active number of flavors in the VFNS. 
This extension provides  additional flexibility in the region of the heavy quark thresholds;
however, the  implementation  of Ref.~\cite{Kusina:2013slm} only used a fixed matching scale of $\mu_m = m_H$.
Using xFitter we can improve on this concept by generating PDFs with a variable $\mu_m$ scale.
We illustrate this below and provide example grids at {\tt xFitter.org}.

\figvfnsii
\figNfPDF

The usual PDF can be generalized  to include an $N_F$-dependence~\cite{Kusina:2013slm}:
$f_{i}(x,\mu)\to f_{i}(x,\mu,N_{F})$.
In this approach, the many $N_{F}=\{3,4,5,...\}$ flavor schemes
coexist, and they can be selected by specifying the number of active
flavors $N_F$ along with the other arguments of the PDF.
This concept is represented pictorially in Fig.~\ref{fig:vfns2}.
All  the $N_{F}$  sets of  PDFs are available above the matching scale $\mu_{m}$.
For example, with an $N_F$-dependent PDF, one could simultaneously  analyze
selected data sets with $N_F=4$ and others with $N_F=5$ 
even if they overlap kinematically;
the user has the flexibility (and responsibility) to select $N_F$.

Note in  Fig.~\ref{fig:vfns2} that the various $N_F$ grids are not individual fits but
are related analytically via the flavor threshold matching conditions.
Operationally, they are generated from an initial PDF $f_i(x,\mu_0,N_F=3)$
and $\alpha_S(\mu_0)$ at the starting scale $\mu_0$.
The $N_F=3$ grid is generated by evolving from $\mu_0$  to  $\mu_{max}$.
The $N_F=4$ grid is then generated by matching at $\mu_c$ (which may or may not equal $m_c$),
and evolving up to scale  $\mu_{max}$.
The $N_F=5$  and  $N_F=6$ grids are generated in a similar manner.\footnote{%
Note the $N_F=\{3,4,5,6\}$ grids are stored in separate LHAPDF data files;
they can be combined into an effective $N_F$ dependent PDF as illustrated
in Refs.~\cite{Kusina:2013slm,Clark:2016jgm}.
} %
This process ensures that all the PDFs  $f_i(x,\mu,N_F)$ are analytically related to
the PDF and $\alpha_S$ boundary conditions at  $\mu_0$.

To provide an explicit illustration of the above,  
we have generated a set of PDF grids with a variety of matching scales ($\mu_b$)
for the matching between the schemes with $N_F=4$ and $N_F=5$ active flavours: 
$\mu_{b}=\{1,3,5,10, \infty \}\times  m_b$.
We focus on $\mu_{b}$ as this is the flavor transition most likely to fall within a particular data set. 
For the initial PDF we use the NNLO bottom fit with $\mu_b=1\, m_b$ of Table~\ref{tab:bottomNNLO}, 
and we evolve at NNLO.
The PDFs are fixed such that they all match at the initial evolution scale $\mu_0=1.0$~GeV
with the  same value of $\alpha_S(\mu_0)= 0.467464$.

This is illustrated in Fig.~\ref{fig:pdfVsQ} where we display the bottom quark and gluon PDFs as a function of $\mu$ in GeV. 
As we evolve up in $\mu$, we explicitly see the transition from $N_F=4$ to $N_F=5$ flavors at each respective $\mu_b$ threshold.
For these particular kinematic values, the discontinuity of the bottom PDF is positive while that of the gluon is negative;
this ensures the momentum sum rule is satisfied. 
Furthermore, we observe  the spread in the bottom PDF at large $\mu$ is broader than
that of Fig.~\ref{fig:bMatch}.
In  Fig.~\ref{fig:pdfVsQ}, while the values of $\alpha_S$ all coincide 
at $\mu_0$, the evolution across the different $\mu_b$ thresholds result in different  $\alpha_S$ values at large $\mu$
scales.
This is in contrast to Fig.~\ref{fig:bMatch}
where  the values of $\alpha_S$ all coincide at the large scale  $\mu=M_Z$.
Additionally, note that the illustration in Fig.~\ref{fig:bMatch} is based on the NNPDF3.0 PDF set while 
Fig.~\ref{fig:pdfVsQ} is based on our fit from  Table~\ref{tab:bottomNNLO}.

Because the $N_F=4$ and $N_F=5$ grids are  available concurrently, 
we can choose to analyze the HERA data  in an $N_F=4$ flavor scheme for arbitrarily large scales,
but simultaneously allow LHC data to be analyzed in a $N_F=5$
flavor scheme throughout the full kinematic region even down to low scales. 

In this illustration, the PDFs revert to  $N_F=4$ below $\mu_b$; however, this is not required.
For example the  $N_F=5$ PDFs could be evolved backwards from $\mu_b$ to provide values at scales $\mu<\mu_b$.
Both APFEL\cite{Bertone:2013vaa} and QCDNUM\cite{Botje:2016wbq,Botje:2010ay}  have this capability.\footnote{%
However, it is generally advisable not to backwards evolve too far in $\mu$ as this can become unstable~\cite{Lomatch:1988uc,Caola:2010cy}.
} %

For bottom at NNLO using the results from Tab.~\ref{tab:bottomNNLO}
for the inclusive data set, we  observe the $\mu_b$ variation is minimal. Thus, a
choice in the range $\mu_{b}\sim [m_{b}, 5 m_b]$ yields a
$\Delta \chi^2 \leq (1457-1453)\sim 4$ units out of $\sim 1450$.
This minimal $\chi^2$ dependence  means we can shift the $\mu_b$ matching scale
if, for example, we want to avoid a $N_F$ flavor transition in a specific kinematic region. 
While these results should be checked with additional data sets, the
insensitivity to $\mu_{b}$, especially at NNLO, is an important result
as the ability to displace the $N_{F}=4$ and $N_{F}=5$ transition 
can be beneficial when this threshold comes in the middle of a data set.

Combined with the variable heavy quark threshold,
the $N_F$ dependent PDFs provide additional flexibility
to analyze multiple data sets in the optimal theoretical context.

\section{Conclusions}
\label{sec:conclusion}

In this study we have examined the impact of the heavy flavor matching
scales $\mu_m$ on a PDF fit to the combined HERA data set.

The choice of $\mu_m$ allows us to avoid delicate cancellations in the
region $\mu_m\sim m_H$ as illustrated in Fig.~\ref{fig:acot}.
Additionally, the discontinuities associated with the $N_F=4$ to
$N_F=5$ transition can be shifted so that these discontinuities do not
lie in the middle of a specific data set.

Using {\tt xFitter} and {\tt APFEL} to study the $\mu_m$ dependence
of a global PDF fit to the HERA data,
we can extract the following general features.  For the charm matching
scale, $\mu_c$, there is a large variation of $\chi^{2}$ at NLO, but
this is significantly reduced at NNLO. In contrast, for the bottom
matching scale, $\mu_b$, there is a relatively small variation of
$\chi^{2}$ at both NLO and NNLO.

These observations can be useful when performing fits. While charm has
a larger $\chi^{2}$ variation (especially at NLO), the charm quark
mass $m_c\sim 1.45$~GeV lies in a region which is generally excluded
by cuts in $Q^2$ and/or $W^2$.

On the contrary, the $\chi^{2}$ variation for the bottom quark is
relatively small at both NLO and NNLO. Since the bottom quark mass
$m_b\sim 4.5$~GeV is in a region where there is abundance of precision
HERA data, this flexibility allows us to shift the heavy flavor
threshold (and the requisite discontinuities) away from any particular
data set. Functionally, this means that we can analyze the HERA data
using an $N_F=4$ flavor scheme up to relatively large $\mu$ scales,
and then perform the appropriate NNLO matching (with the associated
constants and log terms) so that we can analyze the high-scale LHC
data in the $N_F=5$ or even $N_F=6$ scheme.

These variable heavy flavor matching scales $\mu_m$ allow us to
generalize the transition between a FFNS and a VFNS, and provides a
theoretical ``laboratory'' which can quantitatively test proposed
implementations. We demonstrated this with the example of the
$N_F$-dependent PDFs. Having the quantitative results for the
$\chi^2$ variation of the $\mu_{c,b}$ scales, one could systematically
evaluate the impact of using different matching
scale choices for the  $f_i(x,\mu,N_F)$.

In conclusion, we find that the ability to vary the heavy flavor
matching scales $\mu_m$, not only provides new insights into the
intricacies of QCD, but also has practical advantages for PDF fits.

%% file: main.bbl
\begin{thebibliography}{10}
\providecommand{\url}[1]{{#1}}
\providecommand{\urlprefix}{URL }
\expandafter\ifx\csname urlstyle\endcsname\relax
  \providecommand{\doi}[1]{DOI \discretionary{}{}{}#1}\else
  \providecommand{\doi}{DOI \discretionary{}{}{}\begingroup
  \urlstyle{rm}\Url}\fi

\bibitem{Aivazis:1990pe}
M.A.G. Aivazis, F.I. Olness, W.K. Tung, Phys. Rev. Lett. \textbf{65}, 2339
  (1990).
\newblock \doi{10.1103/PhysRevLett.65.2339}

\bibitem{Aivazis:1993kh}
M.A.G. Aivazis, F.I. Olness, W.K. Tung, Phys. Rev. \textbf{D50}, 3085 (1994).
\newblock \doi{10.1103/PhysRevD.50.3085}

\bibitem{Aivazis:1993pi}
M.A.G. Aivazis, J.C. Collins, F.I. Olness, W.K. Tung, Phys. Rev. \textbf{D50},
  3102 (1994).
\newblock \doi{10.1103/PhysRevD.50.3102}

\bibitem{Thorne:2000zd}
R.S. Thorne, R.G. Roberts, Eur. Phys. J. \textbf{C19}, 339 (2001).
\newblock \doi{10.1007/s100520100605}

\bibitem{Martin:2010db}
A.D. Martin, W.J. Stirling, R.S. Thorne, G.~Watt, Eur. Phys. J. \textbf{C70},
  51 (2010).
\newblock \doi{10.1140/epjc/s10052-010-1462-8}

\bibitem{Forte:2010ta}
S.~Forte, E.~Laenen, P.~Nason, J.~Rojo, Nucl. Phys. \textbf{B834}, 116 (2010).
\newblock \doi{10.1016/j.nuclphysb.2010.03.014}

\bibitem{Ball:2011mu}
R.D. Ball, V.~Bertone, F.~Cerutti, L.~Del~Debbio, S.~Forte, A.~Guffanti, J.I.
  Latorre, J.~Rojo, M.~Ubiali, Nucl. Phys. \textbf{B849}, 296 (2011).
\newblock \doi{10.1016/j.nuclphysb.2011.03.021}

\bibitem{Alekhin:2009ni}
S.~Alekhin, J.~Bl{\"u}mlein, S.~Klein, S.~Moch, Phys. Rev. \textbf{D81}, 014032
  (2010).
\newblock \doi{10.1103/PhysRevD.81.014032}

\bibitem{Alekhin:2012ig}
S.~Alekhin, J.~Bl{\"u}mlein, S.~Moch, Phys. Rev. \textbf{D86}, 054009 (2012).
\newblock \doi{10.1103/PhysRevD.86.054009}

\bibitem{Alekhin:2013nda}
S.~Alekhin, J.~Bl{\"u}mlein, S.~Moch, Phys. Rev. \textbf{D89}(5), 054028
  (2014).
\newblock \doi{10.1103/PhysRevD.89.054028}

\bibitem{Stavreva:2012bs}
T.~Stavreva, F.I. Olness, I.~Schienbein, T.~Je{\v z}o, A.~Kusina, K.~Kovar\'ik,
  J.Y. Yu, Phys. Rev. \textbf{D85}, 114014 (2012).
\newblock \doi{10.1103/PhysRevD.85.114014}

\bibitem{Qian:1985xp}
S.~Qian, {The CWZ Subtraction Scheme (A New Renormalization Prescription For
  QCD) And Its Application}.
\newblock Ph.D. thesis, IIT, Chicago (1985).
\newblock \urlprefix\url{http://wwwlib.umi.com/dissertations/fullcit?p8517585}

\bibitem{Collins:1986mp}
J.C. Collins, W.K. Tung, Nucl. Phys. \textbf{B278}, 934 (1986).
\newblock \doi{10.1016/0550-3213(86)90425-6}

\bibitem{Buza:1995ie}
M.~Buza, Y.~Matiounine, J.~Smith, R.~Migneron, W.L. van Neerven, Nucl. Phys.
  \textbf{B472}, 611 (1996).
\newblock \doi{10.1016/0550-3213(96)00228-3}

\bibitem{Olive:2016xmw}
C.~Patrignani, et~al., Chin. Phys. \textbf{C40}(10), 100001 (2016).
\newblock \doi{10.1088/1674-1137/40/10/100001}

\bibitem{Alekhin:2014irh}
S.~Alekhin, et~al., Eur. Phys. J. \textbf{C75}(7), 304 (2015).
\newblock \doi{10.1140/epjc/s10052-015-3480-z}

\bibitem{Zenaiev:2016jnq}
O.~Zenaiev, PoS \textbf{DIS2016}, 033 (2016)

\bibitem{Bertone:2013vaa}
V.~Bertone, S.~Carrazza, J.~Rojo, Comput. Phys. Commun. \textbf{185}, 1647
  (2014).
\newblock \doi{10.1016/j.cpc.2014.03.007}

\bibitem{Bertone:2017inprep}
V.~Bertone, A.~Glazov, A.~Mitov, A.~Papanastasious, M.~Ubiali, \textit{In
  preparation}
  \urlprefix\url{https://indico.hep.anl.gov/indico/getFile.py/access?contribId=89&sessionId=28&resId=1&materialId=slides&confId=1161}

\bibitem{Collins:1978wz}
J.C. Collins, F.~Wilczek, A.~Zee, Phys. Rev. \textbf{D18}, 242 (1978).
\newblock \doi{10.1103/PhysRevD.18.242}

\bibitem{Collins:1998rz}
J.C. Collins, Phys. Rev. \textbf{D58}, 094002 (1998).
\newblock \doi{10.1103/PhysRevD.58.094002}

\bibitem{Kusina:2013slm}
A.~Kusina, F.I. Olness, I.~Schienbein, T.~Je{\v z}o, K.~Kovar\'ik, T.~Stavreva,
  J.Y. Yu, Phys. Rev. \textbf{D88}(7), 074032 (2013).
\newblock \doi{10.1103/PhysRevD.88.074032}

\bibitem{Maltoni:2012pa}
F.~Maltoni, G.~Ridolfi, M.~Ubiali, JHEP \textbf{07}, 022 (2012).
\newblock \doi{10.1007/JHEP04(2013)095, 10.1007/JHEP07(2012)022}.
\newblock [Erratum: JHEP04,095(2013)]

\bibitem{Lim:2016wjo}
M.~Lim, F.~Maltoni, G.~Ridolfi, M.~Ubiali, JHEP \textbf{09}, 132 (2016).
\newblock \doi{10.1007/JHEP09(2016)132}

\bibitem{Ball:2013gsa}
R.D. Ball, V.~Bertone, L.~Del~Debbio, S.~Forte, A.~Guffanti, J.~Rojo,
  M.~Ubiali, Phys. Lett. \textbf{B723}, 330 (2013).
\newblock \doi{10.1016/j.physletb.2013.05.019}

\bibitem{Cacciari:1998it}
M.~Cacciari, M.~Greco, P.~Nason, JHEP \textbf{05}, 007 (1998).
\newblock \doi{10.1088/1126-6708/1998/05/007}

\bibitem{Kniehl:2015fla}
B.A. Kniehl, G.~Kramer, I.~Schienbein, H.~Spiesberger, Eur. Phys. J.
  \textbf{C75}(3), 140 (2015).
\newblock \doi{10.1140/epjc/s10052-015-3360-6}

\bibitem{Bonvini:2015pxa}
M.~Bonvini, A.S. Papanastasiou, F.J. Tackmann, JHEP \textbf{11}, 196 (2015).
\newblock \doi{10.1007/JHEP11(2015)196}

\bibitem{Bonvini:2016fgf}
M.~Bonvini, A.S. Papanastasiou, F.J. Tackmann, JHEP \textbf{10}, 053 (2016).
\newblock \doi{10.1007/JHEP10(2016)053}

\bibitem{Ball:2016neh}
R.D. Ball, V.~Bertone, M.~Bonvini, S.~Carrazza, S.~Forte, A.~Guffanti, N.P.
  Hartland, J.~Rojo, L.~Rottoli, Eur. Phys. J. \textbf{C76}(11), 647 (2016).
\newblock \doi{10.1140/epjc/s10052-016-4469-y}

\bibitem{Ball:2015tna}
R.D. Ball, V.~Bertone, M.~Bonvini, S.~Forte, P.~Groth~Merrild, J.~Rojo,
  L.~Rottoli, Phys. Lett. \textbf{B754}, 49 (2016).
\newblock \doi{10.1016/j.physletb.2015.12.077}

\bibitem{Ball:2015dpa}
R.D. Ball, M.~Bonvini, L.~Rottoli, JHEP \textbf{11}, 122 (2015).
\newblock \doi{10.1007/JHEP11(2015)122}

\bibitem{Lyonnet:2015dca}
F.~Lyonnet, A.~Kusina, T.~Je{\v z}o, K.~Kovar\'ik, F.~Olness, I.~Schienbein,
  J.Y. Yu, JHEP \textbf{07}, 141 (2015).
\newblock \doi{10.1007/JHEP07(2015)141}

\bibitem{Abramowicz:2015mha}
H.~Abramowicz, et~al., Eur. Phys. J. \textbf{C75}(12), 580 (2015).
\newblock \doi{10.1140/epjc/s10052-015-3710-4}

\bibitem{Aaron:2009af}
F.D. Aaron, et~al., Eur. Phys. J. \textbf{C65}, 89 (2010).
\newblock \doi{10.1140/epjc/s10052-009-1190-0}

\bibitem{Abramowicz:1900rp}
H.~Abramowicz, et~al., Eur. Phys. J. \textbf{C73}(2), 2311 (2013).
\newblock \doi{10.1140/epjc/s10052-013-2311-3}

\bibitem{Abramowicz:2014zub}
H.~Abramowicz, et~al., JHEP \textbf{09}, 127 (2014).
\newblock \doi{10.1007/JHEP09(2014)127}

\bibitem{Carrazza:2014gfa}
S.~Carrazza, A.~Ferrara, D.~Palazzo, J.~Rojo, J. Phys. \textbf{G42}(5), 057001
  (2015).
\newblock \doi{10.1088/0954-3899/42/5/057001}

\bibitem{Bertone:2016lga}
V.~Bertone, S.~Carrazza, N.P. Hartland, Comput. Phys. Commun. \textbf{212}, 205
  (2017).
\newblock \doi{10.1016/j.cpc.2016.10.006}

\bibitem{Bertone:2016ywq}
V.~Bertone, et~al., JHEP \textbf{08}, 050 (2016).
\newblock \doi{10.1007/JHEP08(2016)050}

\bibitem{James:1975dr}
F.~James, M.~Roos, Comput. Phys. Commun. \textbf{10}, 343 (1975).
\newblock \doi{10.1016/0010-4655(75)90039-9}

\bibitem{Aaron:2012qi}
F.D. Aaron, et~al., JHEP \textbf{09}, 061 (2012).
\newblock \doi{10.1007/JHEP09(2012)061}

\bibitem{Ball:2014uwa}
R.D. Ball, et~al., JHEP \textbf{04}, 040 (2015).
\newblock \doi{10.1007/JHEP04(2015)040}

\bibitem{Bonvini:2016wki}
M.~Bonvini, S.~Marzani, T.~Peraro, Eur. Phys. J. \textbf{C76}(11), 597 (2016).
\newblock \doi{10.1140/epjc/s10052-016-4445-6}

\bibitem{Clark:2016jgm}
D.B. Clark, E.~Godat, F.I. Olness, Comput. Phys. Commun. \textbf{216}, 126
  (2017).
\newblock \doi{10.1016/j.cpc.2017.03.004}

\bibitem{Botje:2016wbq}
M.~Botje, arXiv:1602.08383 [hep-ph]  (2016)

\bibitem{Botje:2010ay}
M.~Botje, Comput. Phys. Commun. \textbf{182}, 490 (2011).
\newblock \doi{10.1016/j.cpc.2010.10.020}

\bibitem{Lomatch:1988uc}
S.~Lomatch, F.I. Olness, J.C. Collins, Nucl. Phys. \textbf{B317}, 617 (1989).
\newblock \doi{10.1016/0550-3213(89)90535-X}

\bibitem{Caola:2010cy}
F.~Caola, S.~Forte, J.~Rojo, Nucl. Phys. \textbf{A854}, 32 (2011).
\newblock \doi{10.1016/j.nuclphysa.2010.08.009}

\end{thebibliography}
